\documentclass[12pt]{article}
\usepackage{amssymb}
\usepackage{amsmath}
\usepackage{graphicx}

\def\unitmatrix{1\!\!1}
\def\hk{hyper-K\"ahler }
\def\Hk{Hyper-K\"ahler }
\def\Ker{\mathop{\mathrm{Ker}}}
\def\R{\mathbb{R}}
\def\C{\mathbb{C}}
\def\H{\mathcal{H}}
\def\be{\begin{eqnarray}}
\def\ee{\end{eqnarray}}

\newcommand{\nn}{\nonumber}

\def\ex#1{\langle#1\rangle}

\def\=:{=\hspace{-.7em}\raisebox{1.1ex}{.}\hspace{.1em}\raisebox{-0.2ex}{.} }

\def\Re{\mathop{\mathrm{Re}}}
\def\Im{\mathop{\mathrm{Im}}}

\usepackage{a4wide}

\makeatletter
\@addtoreset{equation}{section}

\renewcommand{\thefootnote}{\fnsymbol{footnote}}
\makeatother

\begin{document}
\thispagestyle{empty}
\hbox{}
\vspace{-2cm}
\begin{flushright}
TIT/HEP-535 \\
RIKEN-TH-38\\
UT-05-03\\
{\tt hep-th/0503033} \\
March, 2005
\end{flushright}
\vspace{3mm}

\begin{center}
{\LARGE \bf 
Global Structure of 
Moduli Space \\
for 
BPS Walls 
} 
\vspace{7mm}

{\normalsize\bfseries
Minoru~Eto$^1$, 
Youichi~Isozumi$^1$,
Muneto~Nitta$^1$, 
 Keisuke~Ohashi$ ^1$, \\
 Kazutoshi~Ohta $^2$,
 Norisuke~Sakai $^1$
and
Yuji~Tachikawa$^3$}
\footnotetext{
e-mail~addresses: \tt
meto,
isozumi,
nitta,
keisuke@th.phys.titech.ac.jp;
k-ohta@riken.jp;\\
nsakai@th.phys.titech.ac.jp;
yujitach@hep-th.phys.s.u-tokyo.ac.jp
}

\vskip 1em

{ \it $^1$Department of Physics, Tokyo Institute of 
Technology \\
Tokyo 152-8551, JAPAN  \\[2mm]
$^2$
Theoretical Physics
Laboratory\\
 The Institute of Physical and Chemical Research
 (RIKEN)\\
Saitama 351-0198, JAPAN \\[2mm]
$^3$ Department of Physics, University of Tokyo\\
Tokyo 112-0033, JAPAN
 }
 
\vspace{10mm}

{\bf Abstract}\\[5mm]

{\parbox{14cm}{\hspace{5mm}
We study the global structure of the moduli space of BPS
walls in the Higgs branch of supersymmetric theories with
eight supercharges.
We examine the structure in the neighborhood of a
special Lagrangian submanifold $M$, and find that the dimension 
of the moduli space 
can be larger than that naively
suggested by the index theorem,
contrary to previous examples of BPS solitons.
We investigate BPS wall solutions in an explicit example 
of $M$ using Abelian gauge theory.  
Its Higgs branch turns out to contain 
several special Lagrangian submanifolds including $M$.
We show that the total moduli space of BPS walls is 
the union of these submanifolds. 
We also find interesting dynamics
between BPS walls as a byproduct of the analysis.
Namely, mutual repulsion and attraction between BPS walls sometimes
forbid a movement of a wall and lock it in a certain position;
we also find that a pair of walls 
can transmute to another pair of walls with 
different tension after they pass through.

}}
\end{center}
\vfill
\newpage
\setcounter{page}{1}
\setcounter{footnote}{0}
\renewcommand{\thefootnote}{\arabic{footnote}}

\section{Introduction}
\label{INTRO}

Solitons play a key role in understanding non-perturbative 
dynamics in quantum field theories. 
In supersymmetric (SUSY) theories, the 
Bogomol'nyi-Prasad-Sommerfield (BPS) 
saturated solitons preserve a part of 
SUSY \cite{BPS,WittenOlive}, 
and they are important in understanding
the quantum correction to perturbatively
non-renormalized quantities.
BPS solitons in SUSY theories are 
also important in the brane-world 
scenario~\cite{HoravaWitten}--\cite{RandallSundrum}, 
because they are the necessary ingredients to build 
SUSY models on their worldvolume 
with non-singular space-time.

BPS solitons 
usually contain a number of parameters. 
These parameters are called moduli, which provide massless 
degrees of freedom in low-energy effective Lagrangians on 
the worldvolume of the solitons. 
The kinetic term for them 
are nontrivial, and that makes
the moduli spaces for the BPS solitons fascinating 
subjects.
The BPS solitons in SUSY (possibly gauge) theories with 
codimension four, three, two, and one are called 
instantons, monopoles, vortices, and domain walls, respectively. 
The construction of the moduli space for 
instantons are well-established by 
Atiyah, Drinfeld, Hitchin, and Manin~\cite{ADHM} 
and for monopoles by Nahm~\cite{Nahm}. 
These results have found many physical and mathematical 
applications. 
The moduli space of vortices has also been 
constructed recently by using the D-brane 
configurations~\cite{HT,Eto:2004ii}, 
after the discovery for the case of one vortex~\cite{Auzzi:2003fs}.

Somewhat surprisingly, moduli spaces of domain 
walls were relatively unexplored for a long time
although they are the simplest of such solitons.
We have seen that much attention has been paid to 
BPS domain walls in these years \cite{DvaliShifman}--\cite{Ritz:2002fm} 
as a revival of studies in more than 
ten years ago \cite{Cvetic,Abraham:1992vb}.
One of their motivation is the relation to the
brane-world scenario.
We should construct BPS walls in 
SUSY theories with eight supercharges
for application to the SUSY brane-world.
BPS walls and other BPS solitons in SUSY gauge theories 
or SUSY nonlinear sigma models (NLSM)  
with eight supercharges 
have been studied extensively~\cite{Abraham:1992vb}--\cite{BPSwalls-index}.
Models and soliton solutions in the latter can be obtained 
by those in the former by taking the limit of strong gauge coupling, 
and it is often far easier to obtain (exact) solutions in 
the latter than the former.
The SUSY NLSM with eight supercharges should have the hyper-K\"ahler 
target manifold~\cite{Zu}--\cite{SierraTownsend}. 
The method to obtain \hk manifolds by taking  
the strong gauge coupling limit in gauge theories 
is called the hyper-K\"ahler quotient 
construction~\cite{LR}--\cite{ANS}, 
and it can sometimes yield a target manifold of the form of $T^*M$ 
with a K\"ahler manifold $M$ as its special Lagrangian submanifold.

The moduli space of BPS walls was 
considered in a $d=4$, ${\cal N}=1$ 
generalized Wess-Zumino model~\cite{Shifman:1997hg} 
and in $d=4$, ${\cal N}=1$ SUSY QCD~\cite{Acharya:2001dz,Ritz:2002fm}. 
However, for SUSY theories with eight supercharges, 
the moduli space of BPS walls 
was constructed only for $U(1)$ gauge theory~\cite{GTT2,To}
until recently.
The moduli space of BPS walls 
has been determined completely in the case of 
$U(N_c)$ gauge theories with massive $N_f$ $({}> N_c)$ 
fundamental flavors with eight supercharges~\cite{INOS1,INOS2}. 
It has been found that the moduli spaces for various 
topological sectors glue together to form a complex Grassmannian 
$M=Gr_{N_f,N_c} \simeq 
SU(N_f)/[SU(N_c) \times SU(N_f-N_c) \times U(1)]$, 
which is a special Lagrangian 
submanifold
of the Higgs branch $T^*Gr_{N_f,N_c}$ 
of the corresponding massless model. 
This total moduli space is also valid 
for the massive \hk NLSM 
on $T^*Gr_{N_f,N_c}$,  
which can be obtained by taking the strong coupling limit 
of the gauge theories. 
Hence it is very natural to ask whether the total moduli space of 
BPS walls in $T^*M$ is always $M$. 

The domain walls enjoy also several 
interesting properties~\cite{INOS2}. 
For instance, 
there are mutually penetrable pairs of walls which can pass 
through each other by changing the sign of 
the modulus for their relative distance. 
Up to now it has been found that each of these walls maintain 
its individual identity after passing through. 
On the other hand, certain pairs of walls are mutually 
impenetrable and produces a single compressed wall, 
when the modulus for their relative distance is taken to 
minus infinity. 
So far in previous works,
we find that all pairs of walls are impenetrable 
in Abelian gauge theories and in the NLSM resulting from it.

The purpose of this paper is to investigate the global 
structure of the moduli space of BPS walls 
in the Higgs branch of SUSY theories with 
eight supercharges. 
By considering the local structure around a special Lagrangian 
submanifold $M$, we first find that 
the dimension of the moduli space 
can be larger than that naively 
indicated by the index theorem. 
To our knowledge, this sort of phenomenon of additional 
zero modes has not been found in the case of 
other BPS solitons. 
Taking an Abelian gauge theory to obtain a NLSM for 
$T^*M$ as an explicit example, 
we work out BPS wall solutions.  
Several special Lagrangian submanifolds $M_a$'s besides 
$M$ are found to be contained in its Higgs branch. 
We find that the total moduli space of the BPS walls is 
the union of these special Lagrangian submanifolds $M_a$'s, 
some of which share boundary subspaces.
Thus the total moduli space gains
very ample structure contrary to naive expectation.
The \hk quotient itself 
is given by gluing $T^*M_a$'s together.

By considering the physical reason behind the BPS wall moduli 
space, we also observe an interesting mutual 
repulsion and attraction between BPS domain walls. 
We also find a new phenomenon where a pair of walls can pass through 
each other, resulting in a different pair of walls. 
Namely a pair of walls are transmuted 
to another pair of walls with different tension 
while preserving the total tension after 
their encounter.
We recognize that penetrable pairs of walls are not restricted to 
non-Abelian gauge theories (and its associated NLSM). 
The penetrable pairs can also occur 
in gauge theories with rank larger than unity, such as a direct 
product of $U(1)$ factors. 
In particular, it is quite common to have pairs of walls which
transmute through crossing
when there are two or more factors of $U(1)$ 
gauge groups. We will see that
the occurrence of these phenomena is intimately related to the
existence of multiple compact Lagrangian submanifolds,
which is also responsible for the violation of transversality.

Before we move on to the main part,
we would like to comment on the nature of our
analysis.
We only analyze classical aspects of BPS domain walls
in the main part of our paper.  
Due to the high degree of SUSY,
the quantum correction is under better control than those
with less supercharges.
Furthermore, the target spaces we treat
in the latter half of our paper have
as many tri-holomorphic $U(1)$ isometries as possible.
Hence,
they should give us a firm grip on the correction to the \hk metric,
although that for the general Higgs branch is relatively unexplored
compared to that for the Coulomb branch.
Effective theories on the BPS walls, on the other hand, have
only four supercharges, which implies that
there will be much more diverse
interesting non-perturbative phenomena.
We plan to discuss these matters in a future publication.

The organization of the paper is as follows:
We discuss in section \ref{GA} the BPS flow equation in 
$T^*M$, and study the relation between the dimension of the 
moduli space and the index theorem. 
We then move on to study in section \ref{U1NGTH}
the BPS domain walls in
hypertoric manifolds, \i.e.\ the Higgs 
branch of gauged linear sigma models.  There the reader will find 
explicit and concrete examples 
displaying phenomena uncovered in section \ref{GA}.
We will see how several toric K\"ahler manifolds glue 
together to form the total moduli space of BPS walls.
We  also find interesting dynamics among BPS walls,
using the techniques developed in the previous sections.
We conclude the paper in section \ref{sc:discussion} with 
some discussion.
There are two appendices.
Appendix A will serve as a collection of
definitions concerning the Morse-Smale transversality. 
In appendix B, we analyze the index and vanishing theorem
from the viewpoint of gauged linear sigma models.

\section{BPS flows and the index theorem}
\label{GA}

\subsection{BPS flows on general \hk manifolds}
\label{sc:general-hk}
First we give a brief review of the BPS domain wall 
on general massive \hk sigma models. 
We consider theories with eight supercharges, which 
force us to take a \hk manifold as 
the target space of the NLSM \cite{Zu}. 
\Hk manifolds of real dimension $4n$ are Riemannian 
manifolds with the holonomy group reduced to $Sp(n)$.  
In other words, there are three 
complex  structures $I_a$ $(a=1,2,3)$, satisfying 
the 
relation of imaginary quaternions:
\be
I_a^2 = - \unitmatrix &\text{ and }&
I_a I_b = \epsilon_{abc} I_c \ \ \text{when}\ \ a\ne b
\ee
which are covariantly constant with respect to the metric $g$.
Corresponding to these, there exist a triplet of 
symplectic structures (i.e.~closed non-degenerate 2-forms) 
$\omega_a (\cdot,\cdot) = g (I_a \cdot , \cdot)$. 
\Hk manifolds can be thought of complex  manifolds  
by choosing one complex structure, say $I_3$,
out of  $I_{1,2,3}$. 
Let us define complex coordinates $\phi^i$ ($i=1,\cdots,2n$) using 
the chosen complex structure $I_3$. 
Then the symplectic form for $I_3$ can be written in 
the complex coordinates as
$\omega_\R = i g_{ij^*} d\phi^i \wedge d\phi^{*j}$. 
In addition they naturally have a holomorphic symplectic structure, 
made of the rest of the complex structures:  
it is given by the $(2,0)$-form 
$\omega_\C=g_{ik^*}[(I_1){}^{k^*}{}_{j} + i (I_2){}^{k^*}{}_{j}]
d\phi^i\wedge d\phi^j$.

SUSY NLSM with eight supercharges are defined by 
the target \hk manifolds with 
the complex coordinates $\phi^i$ 
as scalar fields.  
In order for the walls to exist, we need a potential term 
in the Lagrangian lifting the vacuum degeneracy. 
The Lagrangian is severely restricted by high degree of 
SUSY, so that the bosonic part of  the Lagrangian
in $d$ space-time dimensions 
should be written as \cite{AF2} 
\be
 {\cal L}_{\rm boson} 
  = g_{ij^*} \partial_{\mu} \phi^i \partial^{\mu} \phi^{*j} 
    + \, \sum_{\alpha} g_{ij^*} K_{\alpha}^i K_{\alpha}^{*j} 
\label{trihol}
\ee
where $K_{\alpha}^i$ ($\alpha = 1, \cdots, 6 -d$) 
generates an isometry preserving the three complex structures.
Such isometries are called tri-holomorphic Killing 
vectors\footnote{
When one couples the \hk sigma model to $N$ vector 
multiplets $(I=1,\ldots,N)$ 
using tri-holomorphic isometries $K^{i,I}$, 
there appear additional ($6-d$)-parameters $q_{\alpha}^I$ 
$(\alpha=1,\ldots,6-d)$ for each vector multiplet.
When we use the vector 
multiplets just as spurions, they give masses to hypermultiplets 
leading to 
the potential term in the  Lagrangian (\ref{trihol}).
Things will become more interesting and complicated when 
one couples the system above to dynamical vector fields. 
That direction of research will be pursued elsewhere.
}.
We absorbed the mass parameters into the Killing  vectors.
This NLSM with potential term is called 
the massive \hk NLSM,
and it can be obtained by a Sherk-Schwarz dimensional 
reduction \cite{ScherkSchwarz} from a massless NLSM 
in six spacetime dimensions \cite{SierraTownsend}. 
In five dimensions ($d=5$), we can choose one isometry 
$K^i$ so that the potential for the hypermultiplet can 
be given by the square of the single Killing vector. 
Although we can have options of 
choosing two (three, four) different Killing vectors to 
obtain a potential term in four (three, two) dimensions,
it is enough to consider one Killing vector 
to construct 1/2 BPS walls. 
Therefore we take a single tri-holomorphic Killing vector 
for the potential throughout this paper. 

{}From the expression \eqref{trihol}, one can see that the SUSY vacua are 
precisely the fixed points of the $U(1)$ action by $K^i$. 
Let us take coordinates $\phi^i$ so that their origin 
$\phi^i=0$ becomes a chosen fixed point. 
Then, we can expand the Killing vector 
as $K^i= q^i{}_j \phi^j+O(\phi^2)$ near each vacuum. 
Choosing suitable local coordinates, one can arrange so 
that $K^i= iq^i \phi^i$ with $q^i$ real (no summation on $i$). 
The eigenvalue $q^i$ gives the charge and the mass of the 
$i$-th scalar field.

We assume that 
there exist at least two vacua to have domain walls.
Energy density of domain walls per unit volume 
perpendicular to $y$ direction can be bounded from below as follows:
\be
{\cal E}&=&
 \int dy \Bigl(g_{ij*}{\partial \phi^i \over \partial y}
 {\partial \phi^{*j} \over \partial y}
 + g_{ij*}K^iK^{*j}\Bigl)\nonumber\\
 &=&\int dy \, g_{ij*}\Bigl({\partial \phi^i \over \partial y}-i K^i\Bigl)
 \Bigl({\partial \phi^{*j} \over \partial y}+i K^{*j}\Bigl)
 +i\int dy \, g_{ij*}\Bigl(K^i {\partial \phi^{*j} \over \partial y}
 -{\partial \phi^{i} \over \partial y}K^{*j}\Bigl)\nonumber\\
&\ge& 
i\int dy \, g_{ij*}\Bigl(K^i {\partial \phi^{*j} \over \partial y}
-{\partial \phi^{i} \over \partial y}K^{*j}\Bigl)
= [D]_{y=-\infty}^{y=+\infty}\label{tension-walls}
\ee
where $D$ is the moment map 
(or the Killing potential)\footnote{
Since the Killing vector is tri-holomorphic, 
the moment map is also a triplet $\vec{D}= (D_1,D_2,D_3)$.
Using $SU(2)_R$ transformation, we can always take
$[\vec{D}]^{y=+\infty}_{y=-\infty}$ to lie along the third direction.
This choice is most convenient since we have selected
$I_3$ as {\em the} complex structure corresponding to the
four supercharges.
} for 
$K^i$ which is defined by $K^i = -i g^{ij^*} \partial_{j^*} D$.
The inequality is saturated \cite{Abraham:1992vb} if 
the BPS equation
\be
\frac { \partial \phi^i}{ \partial y} =  iK^i
\label{BPSeq}
\ee
 is satisfied, and then the energy density is given by
 $[D]^{y=+\infty}_{y=-\infty}$.

Let us analyze the flow near each of the vacuum. 
Take suitable coordinates as before, so that 
$K^i= i q^i \phi^i$ (no summation) and $q^i$ be real. 
The BPS equation near the vacuum is then approximated by 
\be
\frac{\partial \phi^i}{\partial y}= - q^i \phi^i.
\ee 
This means that the wall flows in or out along the 
$i$-th scalar field
according to the sign of the $U(1)$ charge $q^i$. 
As our theories have eight supercharges, 
$q^i$ come in pairs of $+q $ and $ -q$. 
Hence, out of the $4 n$ real dimensions of the tangent space, 
always $2 n$ is of the incoming direction and $2 n$ is of the 
outgoing direction. 
It is known that the BPS equation (\ref{BPSeq})
is precisely the 
Morse flow with 
the Morse  function $D$ \cite{GTT2}.
The Morse index,
which is defined as the number of
outgoing directions of Morse flows,
is always $2 n$ irrespective of the vacuum. 
Apparently, there is not much of geometrical information 
carried by the Morse flow at this level of generality.

Another point to be noted is that \hk manifolds are Ricci-flat, and that
compact Ricci-flat manifolds with trivial $\pi_1$
admit no Killing vectors.
Therefore our \hk manifolds should be non-compact. 
The Morse function $D$ diverges in general if
the Morse flows are going to infinities along the non-compact directions. 
So these flows give infinite wall tension (\ref{tension-walls}) 
and we have to discard them. 
We will find in the next subsection that
the Morse indices will give rich information when they are 
calculated on a compact submanifold of the \hk manifold.

\subsection{BPS flow near special Lagrangian submanifold}
\label{BPSFIT}

A middle dimensional submanifold $M$ 
with
the holomorphic
symplectic form $\omega_\C $ restricted to $M$ being zero, 
$\omega_\C|_M=0$, 
is called a special Lagrangian submanifold. 
In the following we suppose that 
the target \hk manifold contains 
a {\it compact} special Lagrangian submanifold $M$. 
Here $M$ is K\"ahler with respect to non-vanishing 
$\omega_\R|_M$.
Then, the neighborhood of $M$ inside the \hk manifold 
can be identified with $T^*M$.  
We can utilize this 
description to analyze the structure of the flow near $M$. 
The case of $M = \mathbb{C}P^n$ manifold has been studied 
along these lines in Ref.~\cite{GTT2} 
using the Morse theory, 
and one might suppose that the case already analyzed is
sufficiently generic.
In this section, however, we will encounter with  a lot of surprises.

Let us first recall the \hk structure on $T^*M$. 
The manifold $T^*M$ 
has the holomorphic symplectic structure, 
which is 
the necessary 
conditions for any manifold to have a \hk metric. 
Mathematicians have found \cite{metric on T*M:math} 
that there exists a unique \hk metric $g$ on $T^*M$, which
satisfies the condition: 
\begin{list}{$(*)$}{
}
\item $g$ coincides with the given metric on $M$ 
when restricted on it, and
is invariant under the $U(1)$ 
rotation along the cotangent direction.
\end{list}
There is also physical realization of these metrics using 
the projective superspace formalism \cite{metric on T*M:phys}.
It should  be noted that the metric may only 
be defined on some neighborhood of the zero section and 
that the geodesic length to infinity along the cotangent 
direction may be finite.
This remark will be important
when we study the global structure of the flow in 
section \ref{U1NGTH}.
We focus in this section in the neighborhood of the zero section.
\setlength{\parskip}{2mm}

We decompose the complex scalars $\phi^i$ $(i=1,\cdots,2n)$
into two sets $(z^i,\tilde z_i)$ so that
$z^i$ ($i=1,\cdots, n$) be the local complex coordinates 
of the base $M$ and $\tilde z_i$ be their canonical  conjugates 
with respect to $\omega_\C$ parametrizing  the cotangent direction.
The holomorphic symplectic form 
can be written in these coordinates 
as $\omega_\C=\omega_1+i\omega_2
=dz^i\wedge d\tilde z_i$.
Let $k^i$ be a holomorphic Killing vector of $M$.
The action of $k^i$ on $M$ naturally induces a vector field
$K^i = (k^i, - \tilde z_j \partial_i k^j)$ defined on all of $T^*M$. 
Let us now check that $K^i$ 
preserves the holomorphic symplectic form.
Using the Lie derivatives of the one-forms  
\be
{\cal L}_K dz^i= \partial_j k^i dz^j &\text{and}&
{\cal L}_K d\tilde z_i=
-\partial_i k^j d\tilde z_j-\tilde z_j \partial_i\partial_l k^j dz^l,
\ee 
we have
\be
{\cal L}_K\omega_\C= \partial_j k^i dz^j\wedge d\tilde z_i
+dz^i\wedge(-\partial_i k^j d\tilde z_j-\tilde z_j \partial_i\partial_l k^j dz^l)
=0.
\ee
Secondly, let us show that $K^i$ preserves the K\"ahler form 
$\omega_{\mathbb{R}}$ on $T^*M$.
Consider the integrated flow $g_y:T^*M\to T^*M$ generated by $K^i$:
$dg_y/dy |_{y=0}=K^i$.
{}From the assumption that $k^i$ is a holomorphic isometry of $M$,
we have $g_y^*(\omega_{\mathbb{R}})|_M=\omega_{\mathbb{R}}|_M$.  
Additionally, the action of $K^i$ commutes with the $U(1)$ 
rotation along the
cotangent direction. 
That is, $g_y^*(\omega_{\mathbb{R}})$ 
also satisfies the above condition $(*)$. 
Hence, we have $g_y^*(\omega_{\mathbb{R}})=\omega_{\mathbb{R}}$ 
from the uniqueness.
Summarizing, we have shown 
that $K^i$ is tri-holomorphic on $T^*M$, so that
we can use $K^i$ to give masses to the hypermultiplets.

The BPS equation (\ref{BPSeq}) for walls  then becomes
\begin{align}
 \frac{d z^i}{d y}& = ik^i ,
                      \label{boo}\\
 \frac{d \tilde z_i}{d y}& = -i\tilde z_j \frac{\partial k^j}
  {\partial z^i}\label{bar}
\end{align}when one uses the variables $z^i$ and
$\tilde z_i$. 
Hence, we can first solve the differential equation \eqref{boo} 
for $z$,
and then can solve the equation \eqref{bar} for $\tilde z$  
by plugging in the solution just obtained for $z$.

Let us consider a wall interpolating two vacua
$\alpha$ $(y\to\infty)$ and $\beta$ $(y\to-\infty)$ inside the
base $M$.
Then, the wall satisfies the conditions 
 $\tilde z_i=0$ at $y\to\pm\infty$.
We can think of the study of \eqref{bar} as the study of zero modes
of the operator \begin{equation}
\mathcal{D}_1=\unitmatrix\frac{d}{dy} +i\mathbb{A}\label{on cotangent}
\end{equation}
acting on the functions $\tilde z_i(y)$
where $\unitmatrix$ is the unit matrix
and $\mathbb{A}_i{}^j \equiv \partial k^j/\partial z^i$.

Let us next consider the linearized version of equation \eqref{boo} 
which governs 
the local deformation of the wall profile in the base $M$:
\begin{equation}
 \frac{d\delta z^i}{dy} = i \frac{\partial k^i}{\partial z^j}\delta z^j.
\end{equation}
We set the boundary condition $\delta z^i\to 0$ when $y\to\pm\infty$
so that the solution should describe
normalizable  deformation 
in a fixed topological sector. 
Hence the number of the freedom inside the base space
is given by the number of the zero modes of the operator
\begin{equation}
 \mathcal{D}_2 \equiv 
-\unitmatrix\frac{d}{dy}+i\mathbb{A}^{\rm T}
\label{on deformation}
\end{equation} 
which acts on the space of $\delta z^j$.
Alert reader will have already recognized that the operators
\eqref{on cotangent} and \eqref{on deformation} are adjoint 
to each other, 
and that the boundary conditions placed is precisely those in the 
setup of the index theorem on open spaces~\cite{Callias,APS}.
Thus we immediately obtain the relation 
\begin{equation}
\dim_{\mathbb{R}}\Ker \mathcal{D}_2
-\dim_{\mathbb{R}}\Ker\mathcal{D}_1=
n_{\beta}-n_{\alpha}
\label{index theorem1}
\end{equation} 
where $n_{\alpha,\beta}$ is the Morse index of the flow 
inside  $M$ at the respective vacuum. 
The Morse index at a vacuum is defined as the 
dimension of the outgoing direction of the flow.
It is given explicitly as the number of positive eigenvalues
of the matrix $i\mathbb{A}$ at the vacuum.

Note that the Morse index of the flow on all of $T^*M$ 
is always half of the total dimensions. 
Therefore the above index theorem (\ref{index theorem1}),
if we consider the flow on the
entire target manifold $T^*M$,  
always gives null result trivially\footnote{
This is a generic result applicable not only to 
NLSM, but also to gauged linear sigma models at finite 
gauge coupling, as shown in our Appendix \ref{sc:vanishing-th}.
}. Hence we need to take 
the Morse index of the flow restricted inside the base.

What is the physical implication of the relation 
\eqref{index theorem1}? 
Recall the arguments in the previous subsection. 
The eigenvalues of the matrix 
$\mathbb{A}_i{}^j =\partial k^j/\partial z^i$ at the 
vacuum control the masses of the hypermultiplets there, 
and these in turn control the direction of the flow.
If we restrict attention to the flow inside the base space 
only and denote the total dimensions of the base space by 
$n$, the stable manifold $S(\alpha)$ flowing into the 
vacuum $\alpha$ is of dimension $n-n_\alpha$ and the 
unstable manifold $U(\beta)$ flowing out of the vacuum 
$\beta$ is of dimension $n_\beta$. 
Hence, the BPS flow to $\alpha$ from $\beta$ na\"\i vely 
will form a family of dimension $n_{\beta}-n_{\alpha}$, 
that is,  if we assume 
the stable and the unstable manifolds intersect
transversally\footnote{
We summarize the mathematical definition surrounding 
transversality in Appendix \ref{MSC2}.}.
This is precisely the number appearing in the right hand 
side of the relation \eqref{index theorem1}.
The true dimension of the deformations of the BPS flow 
is, on the other hand,
the first term $\dim\Ker \mathcal{D}_2$  in the left hand side.
These arguments combined tell us that 
$\dim\Ker \mathcal{D}_1$, which counts the freedom of 
deformation in the cotangent direction, 
measures the non-transversality of the flow! 

The index theorem used above is none other than the prototypical 
one which counts the fermionic zero modes in the instanton 
background in $S^3\times \R_y$. 
In that case, $\mathcal{D}_{1(2)}$ corresponds to the 
Dirac operator acting on spinors of positive (negative) 
chirality, and $n_{\alpha, \beta}$ corresponds to the winding 
number of the gauge field in the vacuum at $y=\infty$ 
$(y=-\infty)$. 
The index theorem has been used also 
to count the dimension of BPS domain walls 
in Abelian gauge theories \cite{keith} 
and in non-Abelian gauge theories \cite{BPSwalls-index}. 
There is one crucial difference in the application of 
index theorems to the wall system considered in this paper
 from the cases previously considered. 
For wall systems and instantons to date,
if they are BPS (or in other words they are self-dual 
for the case of instantons), there is a vanishing theorem 
which guarantees the absence of zero modes 
for $\mathcal D_1$. 
Hence, the number of fermionic zero modes  around 
BPS instantons was precisely proportional to the 
instanton number. 
However in our case, there is 
no such vanishing theorem available. 
This means that we can have excess zero modes 
even for BPS walls, and that the dimension of moduli 
of BPS walls may be bigger than the difference of 
indices of the vacua connected. 
We will see an explicit example shortly, and there we 
will also find that this is rather a generic 
phenomenon for BPS walls.

This generalness  necessitates another comment on the 
non-transversality.
When one considers Morse flows in general (without SUSY), 
one can perturb the Morse function a bit to make the flow 
transversal. 
What distinguishes our situation from such generic cases 
is that SUSY forces the flow to be of specific type, 
so that we cannot take a generic perturbation. 
As a result, we are forced to have the non-transversal 
situation\footnote{
It has been already known that the holomorphic isometry 
often violates the transversality, see e.g.\ \cite{BB}. 
}.

When the manifold $M$ is toric, that is, when it can be 
realized using a gauged linear sigma model with four SUSY,
we can obtain $T^*M$ as an open subset of a toric 
\hk manifold, that is, we can realize $T^*M$ as a subset 
of the Higgs branch of a gauged linear sigma model 
with eight SUSY. 
We can study in more detail the reason for the 
non-transversality of the flow in those cases.
Before doing that, however, let us see some typical examples 
exhibiting the somewhat abstract argument above.

\subsection{Example: $T^*\mathbb{C}P^2$}

Let us first study the wall on
the cotangent bundle of a projective space, $T^*\mathbb{C}P^2$.
This system has been analyzed by many groups e.g.\ \cite{GTT2},
and can be understood as a simple example of walls on 
the Grassmannians \cite{INOS1,INOS2}.

$\mathbb{C}P^2$ can be composed from three patches $U_i\sim\mathbb{C}^2$
parametrized by $(v_i,w_i)$,
($i=1,2,3$):
\begin{align}
 (v_1,w_1)&\equiv(v,w), &
 (v_2,w_2)&\equiv(v/w,1/w),&
 (v_3,w_3)&\equiv(1/v,w/v).
\end{align}
The $U(1)^2$ isometry is given by the phase rotation
of $v$ and $w$.
Hence the fixed points, i.e. the vacua, 
are the origins $\alpha_i$ of coordinate patches~\cite{ANS}.
We choose the $U(1)$ subgroup given by
$v\to e^{im_1\theta} v$ and $w\to e^{im_2\theta} w$ to give masses
 to the hypermultiplets, where $m_{1,2}$ are taken to be real.
The isometry acts, near each vacuum, as\be
v_i\to e^{im_1^{(i)}\theta}v_i &\text{\ and \ }&
w_i\to e^{im_2^{(i)}\theta}w_i,
\ee and the BPS flow is given by \be
v_i\to e^{-m_1^{(i)} y}v_i&\text{\ and \ }&
w_i\to e^{-m_2^{(i)}y}w_i
\ee for suitable masses $m_{1,2}^{(i)}$, which are linear combinations
of $m_1$ and $m_2$.

Using the discrete symmetry exchanging $\alpha_{1,2,3}$,
we can assume $0<m_2<m_1$ without loss of generality.
Then, the schematic structure of the flow is 
easily read off to be like 
\begin{equation}
\begin{array}{ccc}
\alpha_1[0]\\
\uparrow&\nwarrow\\
\alpha_2[2]&\leftarrow&\alpha_3[4].
\end{array}
\end{equation}
The Morse index, 
which is denoted in a square
bracket attached to each vacuum, 
is given by
twice the number of outgoing arrows from that vacuum.
One can easily check that there is no violation of 
transversality. 
The dimension of the moduli space of walls with no 
cotangent component between vacua $\alpha_i$ and 
$\alpha_j$ is given by twice the number of sequences 
of arrows connecting them. 

\subsection{Another example: $T^* F_n$}
\label{Fn}
Let us next consider a massive \hk sigma model with the 
target $T^*F_n$. 
Here, $F_n$ is a complex two-dimensional,
real four dimensional surface which is called Hirzebruch 
surface and is defined as the total space of some 
$\mathbb{C}P^1$ fibration over $\mathbb{C}P^1$.
$n$ is an integer describing how the $\mathbb{C}P^1$ is 
fibered on the base. 
We can take $n$ to be positive without loss of generality 
because $F_n\simeq F_{-n}$.
An explicit description of $F_n$ can be given by means of 
four patches
with coordinates $(v_i,w_i)\in U_i\sim \C^2$, $i=1,2,3,4$:
\begin{align}
 (v_1,w_1)&\equiv (v,w), &  (v_2,w_2)&= (1/v,w/v^n), 
\nonumber\\
 (v_3,w_3)&\equiv (v,1/w), &  (v_4,w_4)&= (1/v,v^n/w), 
\label{hirzedef}
\end{align}
where $v$ and $1/v$ parametrize the base $\mathbb{C}P^1$ 
and $w$ and $1/w$ the fiber $\mathbb{C}P^1$. 
These complex surfaces are toric and can be written 
as a vacuum manifold of a gauged linear sigma model 
with four supercharges.  This property will be utilized in 
later sections.

There is a natural isometry on $ F_n$ 
with the standard metric, 
which is given in the first patch by 
\begin{equation}
v\to e^{im_1\theta}v,\qquad
w\to e^{im_2\theta}w.\label{HirzIsom}
\end{equation} 
The action can be 
extended to other patches.
This can be used to give masses to the hypermultiplets. 
The vacua are given by the fixed points of the isometry, 
namely the origins of the four patches $U_i\sim\C^2$, 
which are denoted by $\alpha_i$ respectively.

The BPS flow of the wall corresponding to \eqref{HirzIsom} is
\begin{equation}
v\to e^{-m_1 y}v,\qquad
w\to e^{-m_2 y}w.\label{HirzFlow}
\end{equation}
Let us now consider the flow and the structure of the 
moduli space of walls for some specific $n$, $m_1$, $m_2$.
It can be read off quite easily using the method exemplified 
in the last subsection.
For $m_2<0<m_1$, the flow is schematically given by 
\begin{equation}
\begin{matrix}
\alpha_4[2]&\leftarrow&\alpha_2[4]\\
\downarrow&&\downarrow\\
\alpha_3[0]&\leftarrow&\alpha_1[2]
\end{matrix}.
\end{equation}
One can check that the dimensions of the BPS wall moduli 
in each topological sector have the na\"\i ve ones, that is,
 the difference of the Morse indices.

Next, let us study the case 
for $0<m_2<nm_1$. The flow diagram is changed to 
\begin{equation}
\begin{matrix}
\alpha_4[2]&\leftarrow&\alpha_2[4]\\
\downarrow&&\downarrow\\
\alpha_3[2]&\to&\alpha_1[0]
\end{matrix}.\label{wow}
\end{equation}
Now the vacua $\alpha_3$ and $\alpha_4$ have the same 
Morse index. 
Na\"\i vely, the real dimension of the moduli 
will be the difference of the Morse index, which is zero.
At the same time the BPS wall connecting these two
needs the whole $\mathbb{C}P^1$ to parametrize the profile 
inside the base space $F_n$.
This means, from the discussion in the previous subsection, 
that the BPS wall connecting $\alpha_4$ and $\alpha_3$ is 
allowed to have non-zero expectation values of scalar 
fields corresponding to the cotangent 
direction of $T^*F_n$.
The wall connecting $\alpha_4$ and $\alpha_3$ will show 
very peculiar dynamical behavior in this case.
The analysis can be made much more concrete 
when one uses the gauged linear sigma model 
description. Thus we postpone the study of 
the interesting dynamics hidden in this 
schematic diagram to section \ref{again} 
and let us now move on to investigate 
the walls in the gauged linear sigma model in general.

\section{Walls in hypertoric sigma models}
\label{U1NGTH}

We saw in the previous sections that
the wall moduli space for $T^*F_n$ is bigger than just the 
base space $F_n$, unlike the case for $T^*\mathbb{C}P^n$.
Namely, there exist BPS walls which have some component along
the cotangent direction, thus moving out of the base.
In this section, we realize $T^*F_n$ 
by 
Abelian \hk quotient construction \cite{HKLR},
that is, as the Higgs branch of certain Abelian
gauged linear sigma model.
This viewpoint facilitates the analysis of the
global structure of 
the BPS flow. 
These Abelian \hk quotients are called the
\textit{hypertoric manifolds}\footnote{
These Abelian \hk quotients are also often called 
\textit{toric hyper-k\"ahler}.
However, these are not always toric.
Recall that, in order for a manifold $M$ 
of complex dimension $n$ to be toric,
it has to have $n$ holomorphic isometries.  On the other hand,
an Abelian \hk quotient of complex dimension $2n$ is
guaranteed to have $n$ tri-holomorphic isometries.
This means that it has at least $n$ holomorphic isometries,
but it does not necessarily mean that they have $2n$ of them.
In fact, they are toric if and only if
they are the direct product of several 
ALE manifolds of type A \cite{mathhypertoric}.
We use the word hypertoric in the rest of our paper 
for brevity.
}.
Having gauged linear sigma model description,
we can see  more explicitly and more physically why 
the transversality is violated.

Let us begin with the outline of the 
subsections since this section is rather long.
We first study in detail the Higgs branch of the massless 
model in subsection \ref{WGU(1)GT}.
We will see there that a hypertoric manifold contains 
several compact toric manifolds as special Lagrangian 
submanifolds.
Then followed in subsection \ref{massive} we study
how mass terms given through the tri-holomorphic Killing 
vector determine the discrete vacua.  
We also write down the BPS equations which govern the 
profile of BPS walls.  In subsection \ref{gauged1} and 
\ref{WLTFNLSM}, we will study explicit examples of such 
walls.
These exercises tell us that the total wall moduli space 
are union of the compact special Lagrangian submanifolds,  
glued with each other at precisely the points where 
transversality of the flow may be violated.
This is summarized in subsection \ref{summary}.
We will also see that the existence of multiple 
compact special Lagrangian submanifolds causes 
not only the possible violation of transversality, 
but also interesting dynamics of walls. 
One example is the mutual repulsion and attraction between 
walls treated in subsection \ref{again} and another is 
the transmutation of a pair of walls when they encounter 
with each other, in subsection \ref{Transm}.

\subsection{Abelian gauge theories with charged hypermultiplets}
\label{WGU(1)GT}

Let us first review the construction of hypertoric 
manifolds using physical parlance. 
It is a direct extension to eight supercharges of the 
construction of toric manifolds using Abelian gauge 
theories with four supercharges~\cite{Witten:1993yc}.
The structure of these manifolds is discussed 
in \cite{Ooguri} in relation to the mirror symmetry in three 
dimensions, and in ~\cite{GGPT,Portugues} in relation to
brane solution in supergravity.
Readers who prefer mathematical exposition can 
consult \cite{mathhypertoric}.

Let us consider a $U(1)^N$ gauge theory with 
$N_F$ hypermultiplets with charge matrix $q_I^A$, where $I$ 
and $A$ are the color and flavor indices, respectively. 
Let us denote the field strength and the scalar field 
of the $I$-th multiplet by $F_{MN}^I$ and $\Sigma^I$.
If we denote the $A$-th hypermultiplet with the
$SU(2)_R$ doublet index $i$ by $H^{i}_A$,
the gauge group acts on the hypermultiplet via
\begin{equation}
\delta_\Lambda H^i_A=i \Lambda^I q^A_I H^{i}_A
\label{gaugeaction}
\end{equation}
where $\Lambda^I$ is the gauge parameter for 
the $I$-th $U(1)$. 
The bosonic part of the  Lagrangian is then
\begin{equation}
{\cal L} = -\frac{1}{4g_I^2}(F_{MN}^I)^2 
+ \frac{1}{2g_I^2}(\partial_M \Sigma^I)^2 
+|{\cal D}_M H^{i}_A|^2 -V,
\end{equation} 
where $\mathcal{D}_M H^{i}_A=(\partial_M +iW_M^I q^A_I) H_A^i$
is the covariant derivative. 
The potential $V$ is given by 
\be
V=|(q_I^A \Sigma^I -m^A)H^{i}_A|^2 
+\frac{1}{2g_I^2}(Y^{Ia})^2, 
\ee 
where $m^A \in {\mathbb R}$ is the mass of the $A$-th 
hypermultiplet and 
the auxiliary fields $Y^{Ia}$ are given by their 
equations of motion to be
\be
Y^{Ia}
={g_I^2}\left( c^{Ia}
- \sum_A (H^{i}_A){}^\dagger (\sigma^a)^i{}_j q_I^A H^{j}_A
\right). 
\ee 

In this subsection we consider only the massless case, 
$m^A=0$. 
Let us redefine notations for hypermultiplet scalars in 
order to make four of supercharges apparent as
\be
H^{i=1}_A=h_A, \ \ H^{i=2}_A=\tilde{h}^A{}^\dagger,
\ee
where $h_A$ and $\tilde h^A$ are chiral fields.
It is useful to define the following quantities in examining 
wall solutions and their properties:
\be
\mu_A\equiv h^{A\dagger}h_A -\tilde{h}^A\tilde{h}_A^\dagger, \ \ 
\nu_A \equiv \tilde{h}^A h_A, 
\ee
where index $A$ is not summed. 
By using these notations, the flatness conditions 
determining supersymmetric vacua are
\be
 q_J^A\Sigma^I h_A&=&0, \\
 q_J^A\Sigma^I \tilde h^A&=&0, \\
 q_I^A \mu_A &=&c_I, \label{constraint}\\
 q_I^A \nu_A &=&0.
\ee
Here we have taken the Fayet--Iliopoulos (FI) parameters 
all parallel in $SU(2)_R$ space.
We leave the discussion for non-parallel cases to later works.

Let us concentrate on the Higgs branch, i.e. we take 
$\Sigma^I=0$.
This condition is obtained automatically once one considers 
a generic combination of 
FI parameters $c_I$.
Then the imposed conditions are\begin{eqnarray}
 q_I^A \mu_A =c_I,&\ \text{ and }&
 q_I^A \nu_A =0.\label{FD}
\end{eqnarray}
The vacuum manifold $V$ is obtained by dividing this 
by the action of the gauge group \eqref{gaugeaction}. 
$V$ is a hypertoric manifold obtained by 
the Abelian hyper-K\"ahler quotient.

\begin{figure}
\centerline{\includegraphics[width=.45\textwidth]{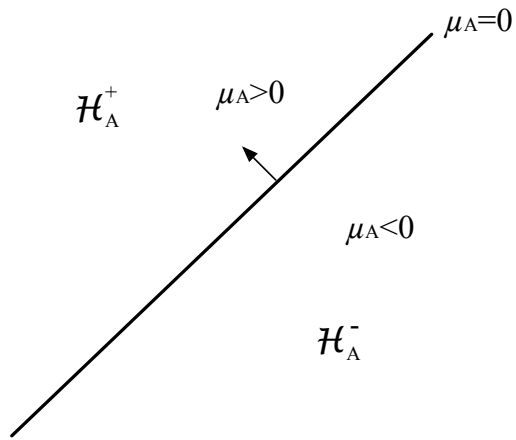}}
\caption{The definition of $\mathcal{H^\pm_A}$.
The arrow indicates in which side  $\mu_A$ is positive.
\label{divide}}
\end{figure}

\begin{figure}
\centerline{\includegraphics[width=.6\textwidth]{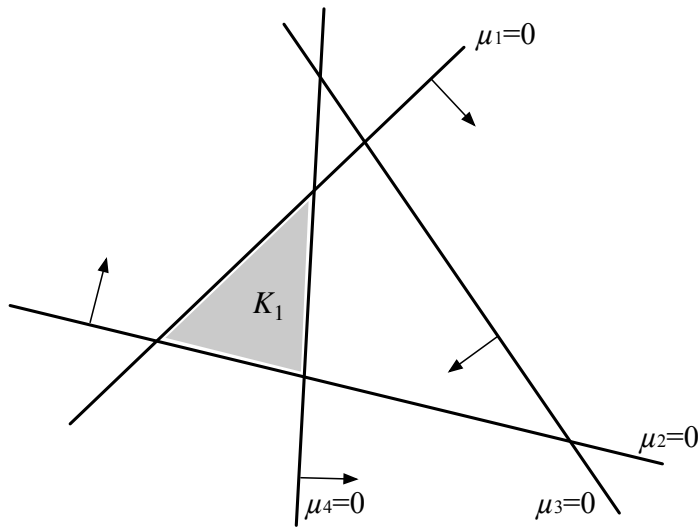}}
\caption{\label{arrangement} An example with $N_F=4, N=2$.}
\end{figure}

Next let us consider the relation to toric manifolds. 
First, recall that the term {\it toric} is just a fancy 
mathematical way of saying that it is a vacuum 
manifold of Abelian gauged linear sigma model with four 
supersymmetries.
Let $K\simeq \mathbb{R}^{N_F-N}$ be the space
spanned by $\mu_A$'s modulo the constraint \eqref{constraint}.
The hyperplane $\mu_A =0$ divides $K$
to half-spaces $\H_A^{+(-)}$ according to the sign of $\mu_A$
as illustrated in Fig.~\ref{divide}.
A small arrow is attached to the hyperplane
in the figure to indicate its orientation.
These hyperplanes divide $K$ to cells $K_a$, $a=1,\ldots,m$.
Let us enumerate them so that $a=1,\ldots, n$ corresponds to 
bounded cells and $a=n+1,\ldots, m$ unbounded ones.
An example with $N_F=4$, $N=2$ is shown in 
Fig.~\ref{arrangement}.

For each cell $K_a$, let us define new variables 
$(\phi_A,\tilde \phi^A)$
based on the relative position of 
the hyperplane $\mu_A$ and $K_a$:\begin{align}
\hat q_I^A&\equiv +q_I^A,&(\phi_A,\tilde\phi^A)&\equiv(h_A,\tilde h^A)
& \text{if\quad }& \H_A^+ \cap K_a \neq \emptyset, \\
\hat q_I^A&\equiv -q_I^A,&(\phi_A,\tilde\phi^A)&\equiv(\tilde h^A, h_A)
&\text{if\quad }&\H_A^- \cap K_a \neq \emptyset.
\end{align}
An illustrative example is depicted in Fig.~\ref{arrangement}.
The shaded region $K$ intersects with $\mathcal{H^+}_1$, 
$\mathcal{H^+}_2$, $\mathcal{H^+}_3$ and $\mathcal{H^-}_4$,
thus, $\phi_A=(h_1,h_2,h_3,\tilde h^4{}^\dagger)$.

Consider a subspace $M_a$ of the hypertoric $V$ defined by 
$\tilde\phi_A=0$.
Together with the vacuum conditions \eqref{FD},
the equation reduces to
\begin{align}
  \sum_A  \hat q_I^A\phi_A\phi_A^\dagger = c_I
\end{align} 
and the gauge equivalence is defined by 
\begin{equation}
\phi_A\sim \exp(i \Lambda^I \hat q_I^A)\phi_A\qquad
\text{(no summation on $A$)}.
\end{equation}
Each subspace $M_a$ 
is a toric manifold defined by the charge matrix 
$\hat q_I^A$ and the FI parameters $c_I$.

Let us determine how $M_a$ is embedded inside $V$.
Consider a point $p=\ex{\phi_A}\in M_a$. 
By expanding \eqref{FD} to the leading order, 
the 
neighborhood of $M_a$ near $p$
is determined by the equations 
\begin{align}
\Re \sum_A\hat q_I^A \delta\phi_A \ex{\phi_A^\dagger} 
&=0, \label{XX}\\
\sum_A\hat q_I^A \delta\tilde\phi^A \ex{\phi_A} 
&=0 \label{YY},
\end{align}  
modulo the gauge variation 
\be
 \delta\phi_A \sim \delta\phi_A
+ i \delta\Lambda^I \hat q_I^A \ex{\phi_A}&
 \text{(no summation on $A$)}.\label{ZZ}
\ee 
Gauge fixing can be naturally attained by extending \eqref{XX}
to 
\begin{equation}
\sum_A\hat q_I^A \delta\phi_A \ex{\phi_A^\dagger} =0. 
\label{WW}
\end{equation}
Eqs.~\eqref{WW} by definition determine 
the tangent space $TM_a\big|_p$ of the manifold $M_a$ at 
$\ex{\phi_A}$.
Comparing \eqref{YY} and \eqref{WW},
the infinitesimal displacement
along $\tilde \phi^A$ direction can be naturally identified 
with $\overline{TM}_a\big|_p\simeq T^*M_a\big|_p$.
As an easy corollary, $M_a$ is a special Lagrangian 
subspace of $V$.
Summarizing, $V$ contains $T^*M_a$  for $a=1,\ldots,m$ 
as subsets.
$M_a$ is compact or noncompact if $K_a$ is bounded or 
unbounded, respectively.
The union of $M_a$'s which is compact is called the 
\textit{core} of $V$.

Before going to the analysis of massive models on hypertoric,
let us study a bit more on the geometry of these manifolds.
Consider a hypertoric $V$ of real dimension $4n$ and notice that
$\mu_A$ and $\nu_A$ modulo relations \eqref{FD} provide
real $3n$ functions to be used as coordinates of $V$,
which are invariant 
by the hypertoric isometries.
Furthermore, points on $V$ with the same values for 
$(\mu_A,\nu_A)$ form
a real $n$ dimensional submanifold.  Thus, it is isomorphic 
to $T^n$.
Summarizing, the hypertoric $V$ can be visualized as a family
of $n$ dimensional tori $T^n$ parametrized by triplet of 
moment maps $(\mu_A,\nu_A)$. 
The subspace $\nu_A=0$ is precisely the union of all $M_a$, 
regardless of whether it is noncompact or not.

\subsection{Massive hypermultiplets}
\label{massive}
Now let us turn on hypermultiplet masses $m^A$ 
to lift the continuous degeneracy of vacua, so that we can 
consider walls interpolating 
the discrete vacua. 
For simplicity and to extract essence, we assume that the 
FI parameters $c_I$ are sufficiently generic 
so that no Coulomb branch remains.
Then, we arrange the masses $m^A$ so that
no hypermultiplet remains massless.
The vacua are now isolated and
the equations determining those vacua are 
\be
(q_I^A\Sigma^I-m^A) h_A&=&0, \label{massive-vac-cond:1}\\
(q_I^A\Sigma^I-m^A) \tilde h^A&=&0, \label{massive-vac-cond:2}\\
q_I^A \mu_A &=&c_I, \\
q_I^A \nu_A &=&0. \label{massive-vac-cond}
\ee

Let us determine vacua by solving these conditions. 
If we consider $N_{\rm F}$ equations $q_I^A\Sigma^I-m^A=0$ 
which define hyperplanes $\sigma ^A$, 
only $N$ equations within them can be
solved at once with the generic $m^A$.  
Thus the first two equations
(\ref{massive-vac-cond:1}) and (\ref{massive-vac-cond:2}) 
imply that 
($N_{\rm F}-N$)-hypermultiplets must vanish. 
Let $\alpha$ be a set of $N$ flavor indices selected among 
$N_{\rm F}$ flavor indices, that is,  
$\alpha \equiv \{ A_1 A_2 \cdots A_N \}$.
Furthermore, let $(q_\alpha)^{A_J}_I$ be a $N\times N$ 
submatrix of 
$q_I^A$  with $\{ A_J \}\in \alpha$. 
If $q_\alpha$ is invertible, i.e.\ 
$\det (q_\alpha)^{A_J}_I \neq 0$, 
the equations for the supersymmetric vacuum can be solved 
as follows:
\be
\langle \Sigma^I \rangle_\alpha  
&=& (q_\alpha^{-1})_I^{A_J}m_{A_J}, \\
\langle \mu_{A_I} \rangle_\alpha 
&=& (q_\alpha^{-1})_J^{A_I}c^J, \\
\langle \nu_{A_I} \rangle_\alpha 
&=&0,
\ee
and we set that all $(h_A,\tilde h^A)$ vanish for 
$\{ A\}\not \in \alpha$. 
The assumptions of generic $c_I$ and $m^A$ imply that 
there are no other kinds of solutions.
The condition 
${\rm det}(q_\alpha)\not = 0$ indicates that each plane within $N$
hyperplanes defined by $\sigma ^{A_I}$ intersects with all the 
other ($N-1$)-hyperplanes. 
Therefore, the above isolated vacua labeled by $\alpha$ 
are given by the intersecting points of 
$N$ hyperplanes in the space of $\Sigma^I$'s.  
The last equation means that either $h_A$ or $\tilde{h}^A$ 
must be zero for each flavor $A$.

These vacua can be also determined from 
the viewpoint of section \ref{GA}.
The set of $m^A$ determines a $U(1)$ subgroup of 
$U(1)^{N_F}$ flavor symmetry,
and this descends to a tri-holomorphic isometry on
the hypertoric manifold. 
Each vacuum corresponds to the fixed point 
of the isometry.
For generic $m^A$,  this fixed point should become a 
fixed point for all the $U(1)^{N_F}$ symmetry.
As the hyperplane $\mu_A=0$ is precisely the 
fixed points of $A$-th $U(1)$ subgroup of $U(1)^{N_F}$,
each vacuum $\alpha$ should be on the intersection of 
($N_F-N$)-hyperplanes 
$\mu_A=0$ ($A\not \in \alpha$) in the space $K$.
Consider the vacuum expectation values 
$(\ex{h_A},\ex{\tilde h^A})$.
They are invariant under $A$-th $U(1)$ for $A\not\in\alpha$. 
They are, however,
not invariant by the $B$-th $U(1)$ for $B\in\alpha$, 
but this can be gauged away
using the $U(1)^N$ symmetry. 
In this sense, the vacuum is labeled by the locking
of flavor and color symmetry.

Let us now move on to study a wall
interpolating two vacua 
labeled by
$\alpha =\{ A_1, A_2, \ldots, A_{N}\}$ and 
$\beta =\{ B_1, B_2, \ldots, B_{N}\}$ respectively.
We would like to determine BPS equations 
and the wall tension by performing the Bogomol'nyi completion 
of wall energy density under the wall ansatz.
The energy density is given by
\be
{\cal E}
&=&
\frac{1}{2g_I^2}(\partial_y \Sigma^I -g_I^2(c_I-q^A_I\mu_A))^2 
+\frac{1}{2g_I^2}| g_I^2 q^A_I \nu_A|^2 \nn\\
&&+|{\cal D}_y h_A+(q_I^A\Sigma^I-m^A)h_A |^2
+|{\cal D}_y \tilde{h}^A -(q_I^A\Sigma^I-m^A)\tilde{h}^A |^2\nn\\
&&+\partial_y\left(c^{I} \Sigma^I
-q_I^A \Sigma^I \mu_A +m^A \mu _A\right).  \label{density}
\ee
The topological charge can be read from the last term to give 
\be
T_{\alpha \leftarrow \beta} 
= \int_{-\infty}^{+\infty} d y \partial_y f,  &\text{\ with\ }&
f \equiv c^I \Sigma^I -q_I^A \Sigma^I \mu_A +m^A \mu _A.\label{Bcompl}
\ee
The BPS equations are obtained from Eq.~(\ref{density}) as 
\be
\partial_y \Sigma^I 
&=& g_I^2(c^I - q_I^A \mu_A),\label{baka}\\
 -2g_I^2 q_I^A \nu_A&=&0,\label{eq:BPS2}\\
\left(\partial_y + q_I^A (\Sigma^I + iW_y^I)\right)h_A 
&=& h_A m^A, \label{hidebu}\\
\left(\partial_y - q_I^A (\Sigma^I + iW_y^I)\right)
\tilde{h}^A 
&=& - \tilde{h}^A m^A \label{abeshi}. 
\ee
When this equations are satisfied, the energy density
becomes equal to $[f]^{y=+\infty}_{y=-\infty}$.

Formal solutions can be found by using the method which we 
have developed in the previous works~\cite{To,INOS1,INOS2}:
First, we define functions $\psi^I(y)$ by the relation
\be
\Sigma^I (y) + iW^I_y(y) &=& \partial_y \psi^I (y).\label{aho}
\ee 
Let us note that $\psi^I$ is defined up to an additive 
constant. 
The hypermultiplets can be expressed using $\psi^I$
by solving (\ref{eq:BPS2}), \eqref{hidebu}, and \eqref{abeshi}:
\be
h_A (y)&=& h_{0A} e^{-q_I^A \psi^I +m^A y},
\label{hyper-sol1} 
\\
\tilde{h}^A (y) &=& \tilde{h}_{0}^A e^{+q_I^A \psi^I-m^A y},
\label{hyper-sol2}
\\
 \nu_A &=&
 \tilde{h}_{0}^A h_{0A} =0.\label{xxx}
\ee
Here, $h_{0A}$ and $\tilde{h}_{0}^A$ are 
vectors with $N_F$ elements.
We call these vectors as moduli matrices in the rest of 
the paper, following the terminology in 
Refs.~\cite{INOS1,INOS2}. 
Eq.~(\ref{xxx}) follows from the boundary conditions 
$\nu_A=0$ at $y\rightarrow \pm \infty$. 
Hence, either $h_A$ or $\tilde{h}^A$ must vanish 
identically for each flavor $A$. 
Thus we find that $\mu_A$ can not flip the sign in a wall configuration.
This fact means that there is no BPS walls which interpolate 
from one vacuum with non-vanishing 
vacuum expectation values (VEV) of 
$h_A$ to another vacuum with non-vanishing VEV of $\tilde{h}^A$. 
That is, BPS walls cannot cross hyperplanes $\mu _A=0$.
Based on these considerations, we can identify sets of vacua 
which can be interpolated  with each other by wall solutions.

Finally, 
$\psi$ itself is solved by plugging \eqref{aho} into \eqref{baka}:
\be
\frac{1}{g_I^2}\partial_y^2 {\rm Re}(\psi^I)
 &=& c^I - q_I^A \mu_A (y) \nn\\
 &=& c^I - q_I^A(|h_{0A}|^2 e^{-2q_J^A {\rm Re}(\psi^J) + 2m^Ay} 
   - |\tilde{h}_{0}^A|^2 e^{+2q_J^A {\rm Re}(\psi^J) - 2m^Ay} 
).\label{kasu}
\ee
Once this second order nonlinear differential equation \eqref{kasu}
is solved for given moduli matrices $h_{0A}$ and $\tilde{h}_{0}^A$, 
all fields can be determined from 
Eqs.~\eqref{hyper-sol1}--\eqref{xxx}. 
Eqs.~(\ref{hyper-sol1}) and (\ref{hyper-sol2}) show that the 
additive ambiguity of $\psi^I$ results in the 
$GL(1,{\mathbb{C}})^N$ equivalence on 
the moduli matrices 
\begin{eqnarray}
 (h_{0A},\tilde h_{0}^A)\simeq 
\left((\lambda _I) ^{q_I^A}h_{0A},
\,(\lambda _I) ^{-q_I^A}\tilde h_{0}^A\right),
\qquad \lambda _I\in \mathbb{C}^*\equiv \mathbb{C}-\{0\}. 
\label{eq:world-vol-sym}
\end{eqnarray}
Thus the equivalence class of the moduli matrices defined 
by Eq.~(\ref{eq:world-vol-sym}) 
describes the moduli spaces of walls,
which are submanifolds of the \hk manifold $V$. 
Note that we eventually take the strong gauge coupling limit 
$g_I\to \infty$ since we are now interested 
only in the Higgs branch
which we constructed using the \hk quotient. 
Therefore the differential equation (\ref{kasu}) reduces to
mere algebraic equations
\begin{equation}
 c_I = q_I^A \mu_A (y)  
 \label{constraints_profile}
\end{equation} 
throughout the wall profile.
Combined with a trivial equation $0 = q_I^A\nu_A (y)$, 
flows defined by the BPS equations can be interpreted as 
Morse flows on the \hk manifold in the strong coupling limit. 
The Morse function for those flows is  
the function $f$ in the Bogomol'nyi completion
\eqref{Bcompl} 
\be
 f &=& c^I \Sigma^I
 -(q_I^A \Sigma^I-m^A)\mu_A=m^A\mu_A.
\ee  
In the following, we will mainly
treat cases in the strong coupling limit 
for simplicity.

\subsection{Walls in gauged linear sigma model 
of the form
$T^*\mathbb{C}P^2$ }\label{gauged1}
\begin{figure}
\centerline{
\includegraphics[width=.4\textwidth]{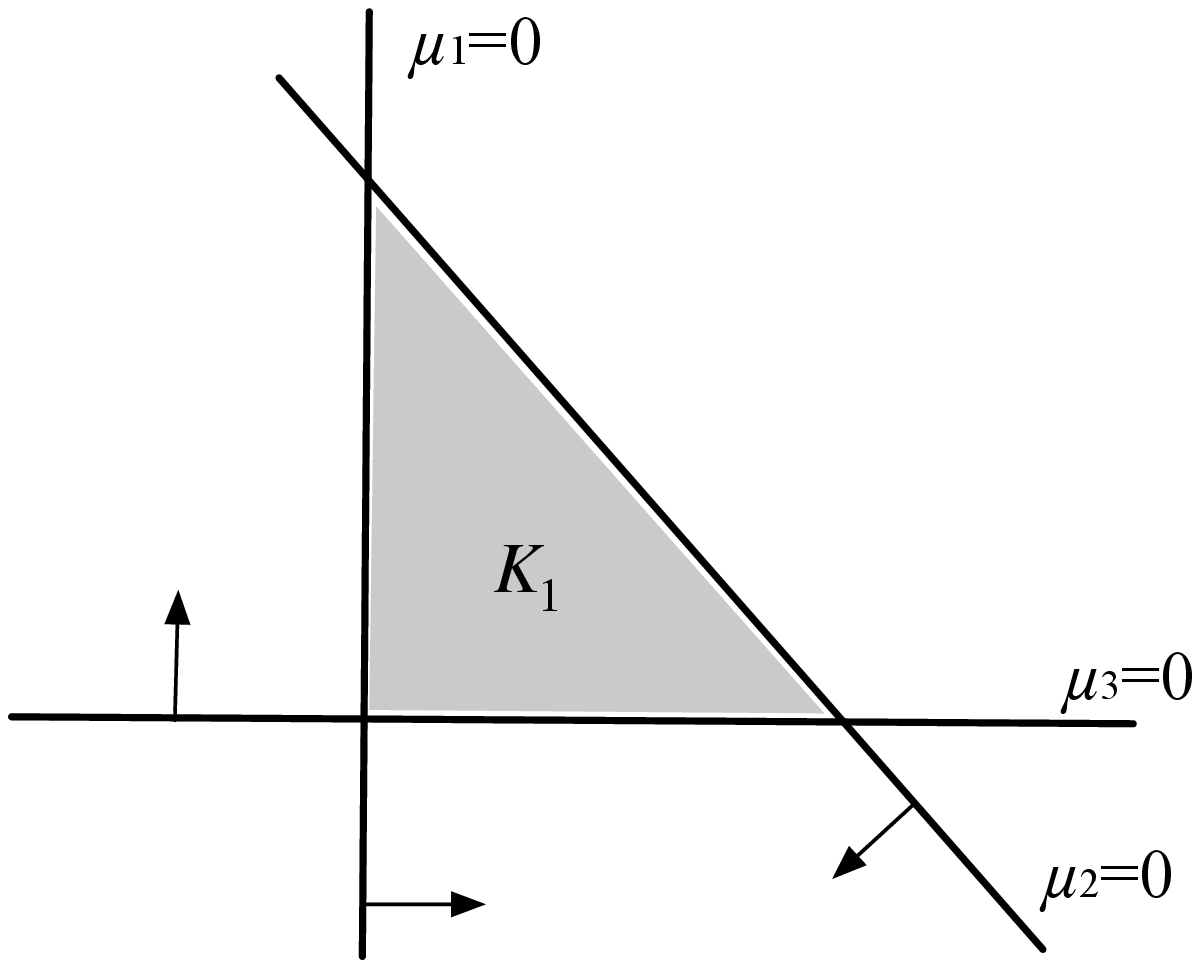}
\includegraphics[width=.5\textwidth]{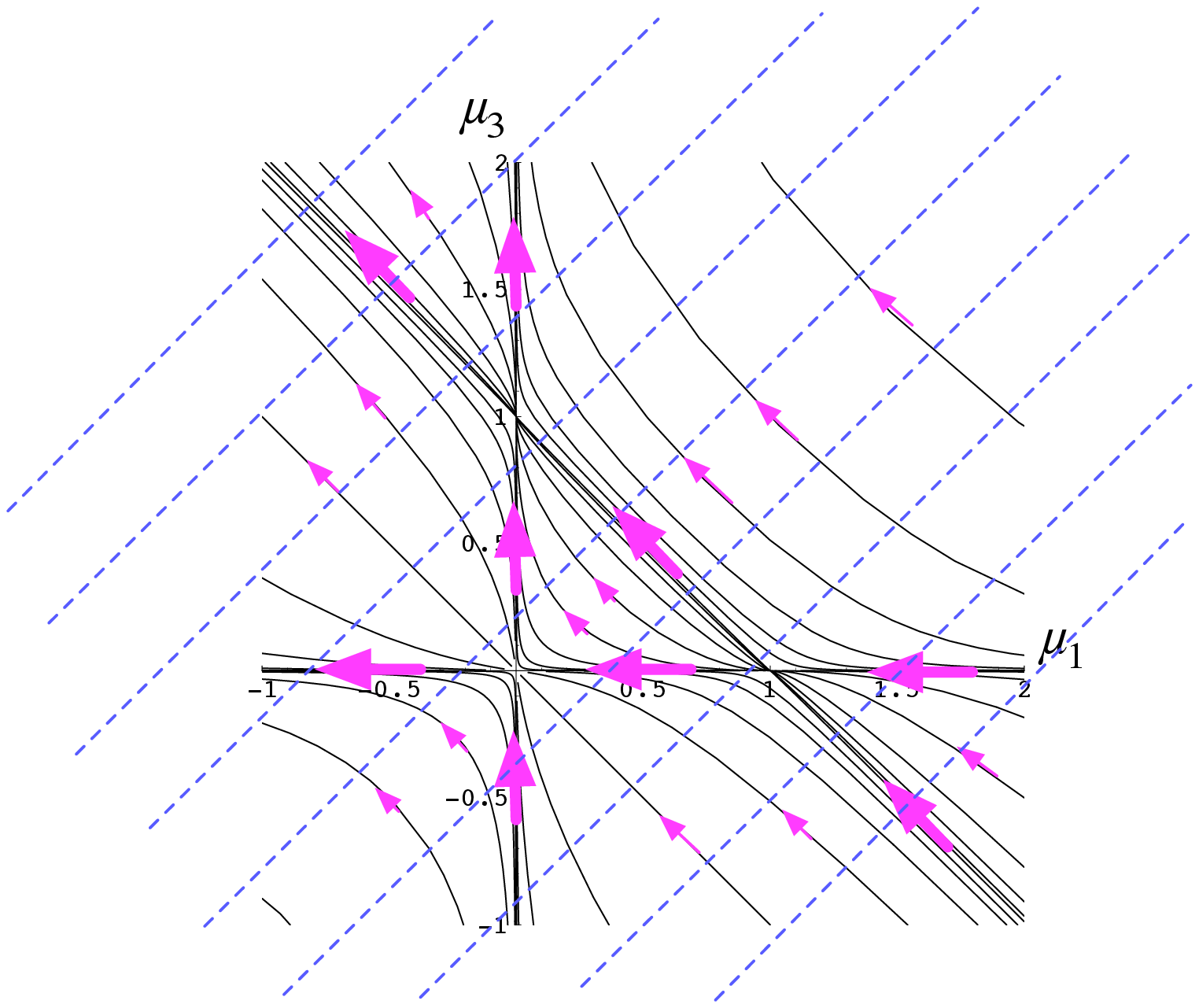}
}
\caption{\label{T*CP2}
Left: Hyperplane arrangement for $T^*\mathbb{C}P^2$;
Right:
Flows in $T^*\mathbb{C}P^2$ with masses $m^A=(1,0,-1)$.
Dashed (blue) lines are the level sets of the Morse function $m^A\mu_A$.
}
\end{figure}
Let us first study the hypertoric realization of $T^*\mathbb{C}P^2$.
$\mathbb{C}P^2$ is specified as a toric manifold by the charge matrix
\begin{equation}
q^A_I=(1,1,1).
\end{equation}
The hypertoric manifold $V$ is determined by the same charge 
matrix. 
The arrangement of  three hyperplanes and the computed flows 
are depicted in Fig.~\ref{T*CP2}. 
{}From this we can see that $V$ contains $\mathbb{C}P^2$ as 
the only compact K\"ahler special Lagrangian submanifold 
and $V=T^*\mathbb{C}P^2$. 
Furthermore, all the vacua of the massive 
$T^*\mathbb{C}P^2$ model reside on $\mathbb{C}P^2$.
Hence, nothing strange happens. 
The moduli space for walls in this case are given by the 
homogeneous coordinates $h_{0A}$ with $\tilde h_0^A=0$ 
under the identification 
\begin{eqnarray}
 (h_{01},\,h_{02},\,h_{03})
\simeq \lambda (h_{01},\,h_{02},\,h_{03}),
\end{eqnarray}
which describe $\mathbb{C}P^2$.
All flows satisfy the transversality. 
We can chase the flow  
starting from a point outside the base $\mathbb{C}P^2$. 
They all go to infinity in the field space, however,
as $y$ goes to $\pm \infty$ as one can see.
Therefore
they do not correspond to walls interpolating between 
two vacua.\footnote{
This statement has been proved in Appendix C of 
Ref.~\cite{INOS2} for more general case of 
$V = T^*Gr_{N_{\rm F},N_{\rm C}}$ 
without taking strong coupling limit.}
What will happen then, when there are vacua outside the 
original base toric manifold?  We will see this in the 
next subsection and how this is related to the violation 
of transversality.

\subsection{Walls in the gauged linear sigma model containing $F_n$ }
\label{WLTFNLSM}

Let us next examine the 
hypertoric manifold containing 
$T^{*} F_n$. 
This model admits walls which 
flow into cotangent directions.
A charge assignment matrix giving this manifold by the quotient construction 
is as follows:
\begin{eqnarray}
q_I^A =\left( 
\begin{array}{cccc}
1& 1& 0& 0\\
0&-n& 1& 1\\
\end{array} 
\right) .
\end{eqnarray} 
Using this charge matrix, we can write down concretely
the flatness conditions 
(\ref{massive-vac-cond:1})-(\ref{massive-vac-cond}) 
for SUSY vacua 
\be
\begin{array}{rclrcl}
 (\Sigma^1 -m_1)h_A&=&0, &
\qquad (\Sigma^1-n\Sigma^2 -m_2)h_A&=&0,\\
 (\Sigma^2 -m_3)h_A&=&0,&
\qquad (\Sigma^2 -m_4)h_A&=&0,
\end{array}
\ee
together with 
the same equations for $\tilde{h}_A$, and 
\be
\begin{array}{rclrcl}
 \mu_1 + \mu_2&=&c_1, &\qquad
 -n\mu_2 + \mu_3 + \mu_4&=&c_2 ,\\
 \nu_1 + \nu_2&=&0, &
 \qquad -n\nu_2 + \nu_3 + \nu_4&=&0.
\label{vacFn}
\end{array}
\ee
Isolated SUSY 
vacua are determined by these equations, 
which have five independent solutions. 
We present these solutions, which are labeled 
from $\alpha_1$ to $\alpha_5$, in Table \ref{table}.

\begin{figure}
\centerline{\includegraphics[width=.6\textwidth]{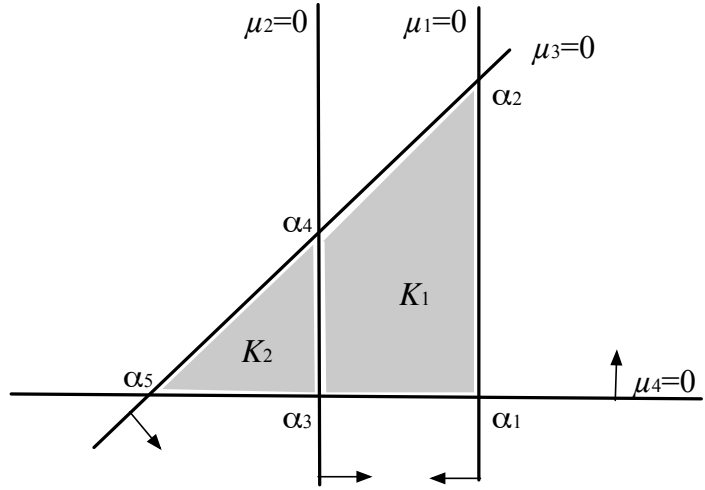}}
\caption{\label{arrangefn}Hyperplane arrangement for the  hypertoric
which contains $T^*F_n$.}
\label{toricFn}
\end{figure}

\begin{table}
\begin{center}
\begin{tabular}{|c||c|c||c|c|c|c|}
\hline
vacua & $\Sigma^1$ & $\Sigma^2$ & $\mu_1$ & $\mu_2$ & $\mu_3$ & $\mu_4$ \\ \hline \hline
$\alpha_1$ & $m_2+nm_3$ & $m_3$ & $0$ & $c_1$ & $c_2 + nc_1$ & $0$ \\
$\alpha_2$ & $m_2+nm_4$ & $m_4$ & $0$ & $c_1$ & $0$ & $c_2 + nc_1$ \\
$\alpha_3$ & $m_1$ & $m_3$ & $c_1$ & $0$ & $c_2$ & $0$ \\
$\alpha_4$ & $m_1$ & $m_4$ & $c_1$ & $0$ & $0$ & $c_2$ \\
$\alpha_5$ & $m_1$ & $(m_1-m_2)/n$ & $c_1 +c_2/n$ & $-c_2/n$ & $0$ & $0$ \\
\hline
\end{tabular} 
\end{center}
\caption{The vacua of massive sigma models which contains $F_n$
in the sigma model target.}
\label{table}
\end{table}
Note that, for any generic choice of the FI parameters, 
there exists at least one vacuum where 
both $h_A$ and $\tilde{h}^A$ have non-zero VEVs.
For simplicity, let us choose the sign of two FI parameters 
to be positive from now on. 
Then, only the vacuum $\alpha _5$ has the non-zero VEV 
of $\tilde{h}^2$.
We show the corresponding
hyperplane arrangement and the placement of vacua in 
Fig.~\ref{arrangefn}.
As depicted there,
there are two compact toric submanifold $M_1$ and $M_2$ corresponding to
the bounded cells $K_1$ and $K_2$.
$M_1$ contains vacua $\alpha_{1,2,3,4}$ and 
$M_2$ contains $\alpha_{3,4,5}$.
In the former $\tilde{h}^A$ vanish 
identically but in the latter this is not the case. 
Since signs of $\mu_A$'s can not change in the wall configuration, 
BPS walls cannot interpolate between the vacuum $\alpha_5$ and 
$\alpha_{1,2}$.
The total moduli space of BPS walls of this model 
is the union of $M_1$ and $M_2$.
The intersection of the two
$M_1 \cap M_2 $  is complex one dimensional.

Homogeneous  and inhomogeneous coordinate representations 
of $M_a$ $(a=1,2)$, can be tabulated as follows: 
Firstly, for $a=1$, homogeneous coordinates
describing $M_1$ are given by
\be
M_1
\ : \ (h_{01}, h_{02}, h_{03}, h_{04}), \ {\rm others}=0,
\ee
with the $GL(1,{\mathbb{C}})^2$ symmetry
\begin{eqnarray}
 (h_{01}, h_{02}, h_{03}, h_{04}) 
 \simeq (\lambda _1h_{01},\, 
 \lambda _1h_{02},\, h_{03},\, h_{04})
 \simeq (h_{01},\, \lambda _2^{-n}h_{02},\, 
 \lambda _2h_{03},\, \lambda _2h_{04}). \label{wv-sym}
\end{eqnarray}
Therefore, the coordinate patches with inhomogeneous coordinates $(v,w)$
can be determined as follows:
\begin{center}
\begin{tabular}{|c||c|c|c|c|}
\hline
patch  & $h_{01}$ & $h_{02}$ & $h_{03}$ & $h_{04}$ \\ \hline 
$U_1$ & $w$ & $1$ & $1$ & $v$ \\
$U_2$ & $w/v^n$ & $1$ & $1/v$ & $1$ \\
$U_3$ & $1$ & $1/w$ & $1$ & $v$ \\
$U_4$ & $1$ & $v^n/w$ & $1/v$ & $1$ \\
\hline
\end{tabular}.
\end{center}
We can see that the transition functions for $M_1$
are precisely the ones for $F_n$ as described in section \ref{Fn}.

Secondly, let us analyze the case $a=2$:
homogeneous coordinates are given by 
exchanging $h_{02}$ for $\tilde{h}_{0}^2$ as
\be
M_2
\ : \ (h_{01}, \tilde{h}_{0}^2, h_{03}, h_{04}), 
\ {\rm others}=0.
\ee 
with the  $GL(1,{\mathbb{C}})^2$ symmetry, 
\begin{eqnarray}
(h_{01}, \tilde{h}_{0}^2, h_{03}, h_{04}) 
\simeq (\lambda _1h_{01},\, 
\lambda _1^{-1}\tilde{h}_{0}^2,\, h_{03},\, h_{04})
\simeq (h_{01},\, \lambda _2^{n}\tilde{h}_{0}^2,\, 
\lambda _2h_{03},\, \lambda _2h_{04}).
\end{eqnarray}
In this manifold $h_{01}$ always takes non-zero values 
because of $\mu _1>0$ and can be fixed 
to one by the first symmetry shown above. Then 
the second symmetry defines $M_2={W\mathbb{ C}P}^2{}_{1,1,n}$, 
which is described by three 
coordinate patches with inhomogeneous coordinates $(u,v)$ as 
follows:
\begin{center} 
\begin{tabular}{|c||c|c|c|c|}
\hline
patch  & $h_{01}$ & $\tilde{h}_{0}^2$ & $h_{03}$ & $h_{04}$ \\ \hline 
$\widetilde{U}_3$ & $1$ & $1/u^n$ & $1$ & $v$ \\
$\widetilde{U}_4$ & $1$ & $1/(uv)^n$ & $1/v$ & $1$ \\
$\widetilde{U}_5$ & $1$ & $1$ & $u$ & $uv$ \\
\hline
\end{tabular} .
\end{center}
Next, let us concentrate on the intersection of two submanifolds where  
$\mu _2=0$, that is, $w\rightarrow \infty $ in $M_1$ and $u\rightarrow \infty $ in $M_2$,
and we find 
\begin{eqnarray}
(h_{01}, 0, h_{03}, h_{04}) 
\simeq (1,\, 0,\, 1,\, v)\simeq (1,\, 0,\, 1/v,\, 1).
\end{eqnarray}
It is thus clear that $M_1 \cap M_2 ={\mathbb{C}P}^1$ 
and the vacua $\alpha _3$ and $\alpha _4$ are on that intersection. 

Therefore, we conclude that the total moduli space 
of walls in this model 
is the union of special Lagrangian submanifolds $F_n$
and $W \mathbb{ C}P^2 {}_{1,1,n} $ with 
${\mathbb{ C}P}^1$ as their intersection.

Now let us write down wall solutions explicitly
and investigate directions of flows. 
By changing mass parameters, flow directions and 
indices at each vacuum change. 
Types of flow can be classified into the following three 
cases:
\be
&&\text{Case I} ; \ 
m_3 > m_4 , \ m_2 +nm_3 > m_1 > m_2 +nm_4, \nn\\
&&\text{Case II} \ ; \ 
m_3 > m_4 , \ m_1 > m_2 +nm_3, \nn\\
&&\text{Case III}\ ; \ m_3 > m_4 , \ m_2 +nm_4 > m_1, \nn
\ee 
and negative thereof.

\subsubsection{Case I}
\label{I}
In the case I, $m_3 > m_4 , \ m_2 +nm_3 > m_1 > m_2 +nm_4$, 
the flow structure is as follows
\be
\left.
\begin{array}{ccccccc}
&&\alpha_4[4,2]& \leftarrow& \alpha_2[4]\\
&\swarrow&\downarrow& & \downarrow\\
\alpha_5[2]&\rightarrow&\alpha_3[0,2]& \rightarrow& \alpha_1[0]\\
\end{array} \right. .
\label{case2}
\ee
We calculated the Morse indices
of the flow inside $M_1$ and $M_2$
and showed the results in the square brackets
after the symbol designating the vacua. 
For $\alpha _3$ and $\alpha _4$ which are contained in both 
$M_1$ and $M_2$,  the second numbers in the square brackets 
denote Morse indices on $M_1=F_n$, and the first numbers 
in them are for $M_2=W\mathbb{C}P^2{}_{1,1,n}$. 
The wall connecting $\alpha_4$ and $\alpha_3$
 breaks transversality from the viewpoint of $F_n$,
because two vacua $\alpha_4$ and $\alpha_3$ have the same 
Morse index. 
{}From the index theorem in Eq.~(\ref{index theorem1}), 
there should be BPS walls 
which flow out along the cotangent direction from $\alpha_4$
and come back from that direction to $\alpha_3$.
This is  precisely the walls depicted in the left half of 
the diagram \eqref{case2}.

Let us explicitly work out walls interpolating between 
vacua $\alpha_2$ and $\alpha_1$.
For simplicity, let us set $(m_1,m_2,m_3,m_4)=(0,0,m,-m)$ 
and $c_1=c_2=c$. 
We fix the gauge for $GL(1,\mathbb{C})^2$ symmetry and take 
the moduli matrix of the topological sector 
$\alpha_2\rightarrow\alpha_1$
to be
\be
(h_0^1,h_0^2,h_0^3,h_0^4)=
 \sqrt{c} \left(a,\,1,\,{1\over \sqrt{2}}e^{-\tau },
  \,{1\over \sqrt{2}}e^{\tau } \right).
\ee
Parameters $a$ and $\tau$ are the moduli.
The configuration corresponding to the above moduli matrix 
contains three walls; 
the first one flows from $\alpha_2$ to near $\alpha_4$, 
the second flows from near $\alpha_4$ to near $\alpha_3$ 
and the third from near $\alpha_3$ to $\alpha_1$. 
We call these walls $W_1$, $W_2$ and $W_3$ respectively as 
in section \ref{Fn}.

Before constructing explicit wall solution 
$\alpha_2\rightarrow \alpha_1$, 
let us estimate wall positions in terms of these moduli 
$a, \ \tau$.
We now assume that three walls
$W_{1,2,3}$ are well 
separated with each other for simplicity. Then 
field configurations are approximately equal to the vacuum 
configuration away from wall centers, since 
Eq.~(\ref{vacFn}) are close to be satisfied.
{}From Table \ref{table}, we find that 
the configuration becomes far from the vacuum 
$\alpha_2$ and $\alpha_1$ ($\alpha_4$ and $\alpha_3$) 
as $\mu_1$ ($\mu_2$) increases. 
So we expect
$W_1$ and $W_3$ 
is situated around the point where the values of $\mu_1$ and $\mu_2$
cross each other, 
namely $\mu_1 \approx \mu_2$. 
To express this condition and construct the wall solution, 
it is useful to introduce the following functions:
\be
 X(y) \equiv e^{-2{\rm Re}\psi ^2}, \ \ 
 Y(y) \equiv e^{-2{\rm Re}\psi ^1},\ \ 
 Z(y) \equiv \cosh{2(m y-{\rm Re}\tau )}.
\ee
$\mu_A$ are simply expressed in terms of $X$, $Y$ and $Z$
\be
\mu_1=c|a|^2Y,\quad 
\mu_2=c\frac{Y}{X^n},\quad
\mu_3+\mu_4=cXZ.
\ee
Then, the condition $\mu_1 \approx \mu_2$ implies 
$|a|^2 =X^{-n}$. 
On the other hand, in the limit that $\mu_1\gg \mu_2$
we obtain $|a|^2\gg X^{-n}\sim Z^n$
by solving Eq.~(\ref{vacFn}). 
Therefore wall positions for $W_1$ and $W_3$ can be roughly 
estimated by the condition $|a|^2\sim Z^n$, which leads to 
the following result
\be
y\sim y_\pm \equiv 
{1\over m}{\rm Re}(\tau )\pm {1\over n}\log{|a|}, \qquad 
{\rm if~} |a|\gg 1. \label{positionA}
\ee

Similarly, the position of the wall $W_2$ for 
$\alpha _4\rightarrow \alpha _3$ 
turns out to be $y\sim y_0\equiv {\rm Re}(\tau )/m$ 
by considering the crossing of $\mu _3$ and $\mu _4$.
Note that the middle wall $W_2$ freezes at the center of the 
outer two walls $W_1$ and $W_3$, 
since there are only two moduli for three constituent walls.
This situation never occurs among walls in 
the case of $T^*\mathbb{C}P^n$, where all flows satisfy the 
transversality.
So we can say that the breaking of the transversality causes 
this result.

Now let us construct these BPS wall solutions explicitly. 
As we take the strong coupling limit  $g^2\rightarrow \infty $, 
we obtain the following algebraic equations from 
Eq.~(\ref{constraints_profile}) 
\begin{eqnarray}
  1&=&Y|a|^2+{Y\over X^n},\quad 
1=-n\,{Y\over X^n}+X Z.
\end{eqnarray}
Eliminating $Y$, we obtain the $n{+}1$-th order equation in $X$
\be
|a|^2ZX^{n+1}-|a|^2X^n+XZ-(n+1)=0.
\ee
For $n=1$,  we can solve the above equation explicitly as 
\begin{eqnarray}
Y&=&{X\over |a|^2X+1},\nn\\
X&=&{1\over 2}\left({1\over Z}-{1\over |a|^2}\right)
+\sqrt{{1\over 4}\left({1\over Z}+{1\over |a|^2}\right)^2
+{1\over |a|^2 Z}}.
\end{eqnarray}
We must plug in these results to moment maps. 
The final result is
\be
\mu _2={c\over |a|^2X+1},&&
 \mu _4={c\over 2}Xe^{-2(m y-\Re\tau )}.\label{mom}
\ee 
We used two linearly independent ones from $\mu_{1,2,3,4}$. 
If one lets the outer two walls going to 
infinities $y\rightarrow \pm \infty$, by taking 
the limit $|a|^2\rightarrow \infty $, 
one finds that $X=Z^{-1}$, $Y=0$ and then
\begin{eqnarray}
 \mu _1=c,\quad \mu _2=0,\quad 
 \mu _3=c{e^{-2(m y-\Re\tau )}\over \cosh{2(m y-{\rm Re}\tau )}},
\quad 
 \mu _4= c{e^{2(m y-\Re\tau )}\over \cosh{2(m y-{\rm Re}\tau )}},
  \label{singlewall} 
\end{eqnarray}
which is just the single-wall solution interpolating 
$\alpha _3$ and $\alpha _4$ 
in the case of ${\mathbb{ C}P}^1$.

\begin{figure}
\begin{center}
\includegraphics[width=15cm,clip]{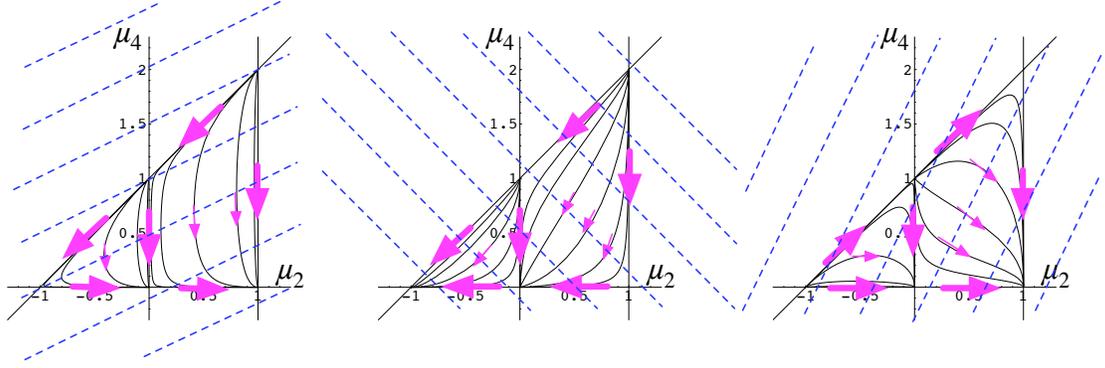}
\end{center}
\caption{Numerically calculated structure of the flow 
in $n=1$ case.
{}From left to the right: case I) $m^A=(0,0,1,-1)$ ;
case II) $m^A=(1,0,0,-1)$ ;
case III) $m^A=(-1,0,1,0)$. 
For all cases $c_I=(1,1)$. 
Dashed (blue) lines designate the contours of constant 
$m^A\mu_A$.
}
\label{numericalflow}
\end{figure}

Next, let us study the walls connecting $\alpha_4$ and $\alpha_3$ 
through $M_2$. 
We express the moduli matrix using $\tau$ as 
\be
 (h^1_0,\tilde h^2_0,h^3_0,h^4_0) 
  = \sqrt{c} \left(b,1,{1\over \sqrt{2}}e^{-\tau },
            {1\over \sqrt{2}}e^{\tau } \right) .
\ee 
We obtain the following two algebraic equations for $X$ and 
$Y$ as before:
\be
 c = cY|b|^2-c{X^n\over Y},&&
 c = n\,c{X^n\over Y}+cX Z.
\ee 
Positions of the two walls are approximated by
\begin{eqnarray}
 y\sim y_\pm \equiv 
{1\over m}{\rm Re}(\tau )\pm {1\over n}\log{|b|},\qquad 
{\rm if~} |b|\gg 1, \label{positionB}
\end{eqnarray}
which result from the condition $\mu _2\sim \mu _3+\mu _4$. 
The explicit solution for $n=1$ is given by 
\be
X&=&{Y\over ZY+1},\nn\\
Y&=&{1\over 2}\left({1\over |b|^2}-\frac1Z\right)
+\sqrt{{1\over 4}\left({1\over |b|^2}-\frac1Z\right)^2
+{2\over |b|^2 Z}}.
\ee 
and  the relation to the moment maps is
\be
\mu _2={-c\over ZY+1},&&
 \mu _4={c\over 2}Xe^{-2(m y-{\Re}\tau )}.
\ee
The structure of the flows is depicted in the left-most 
figure of Fig.~\ref{numericalflow}. 
If one compresses the two walls into a single wall by taking 
$|b|^2\rightarrow 0$, 
one finds that $X=Z^{-1}$ and $Y=\infty $, 
reproducing the result (\ref{singlewall}). 
We thus have found that two moduli spaces 
intersect through ${\mathbb{ C}P}^1$.

\subsubsection{Case II \& III}
\label{IIandIII}
In Case II  
, $m_3 > m_4 , \ m_1 > m_2 +nm_3$, 
the flow structure is as follows
\be
\left.
\begin{array}{ccccccc}
&&\alpha_4[4,2]& \leftarrow& \alpha_2[4]\\
&\swarrow&\downarrow& & \downarrow\\
\alpha_5[0]&\leftarrow&\alpha_3[2,0]& \leftarrow& \alpha_1[2]\\
\end{array} 
\right. 
\label{case1}.
\ee
In the right half of the above diagram, the square
made up of $\alpha_{1,2,3,4}$ denotes the projection of $F_{n}$
by the moment map. 
One can check there are no walls breaking the Morse-Smale
transversality condition.
{}From the general discussions in section \ref{GA},
there should be no walls flowing out to 
and coming back from the cotangent direction.
Indeed, any flow from $\alpha_4$ with arbitrarily little 
cotangent component goes to the vacuum $\alpha_5$ and never 
comes back to $\alpha_3$.

To construct explicit wall solutions, let us take the 
hypermultiplet masses 
\begin{equation}
m^A=(qm, 0, 0, -m), 
\end{equation}
as an illustrative example. 
Using the symmetry (\ref{wv-sym}), 
the moduli matrix can be fixed to the following form with 
the complex moduli parameters $\tau_1, \tau_2$ 
\be
(h_0^1,h_0^2,h_0^3,h_0^4)=
 \sqrt{c} \left(e^{-q\tau_1},\,1,\,1,\,e^{\tau_2} \right).
\label{eq:moduli-matrix-II}
\ee
By a similar analysis at strong gauge coupling as in the 
case I, we find the wall tensions 
and positions as summarized in Table \ref{table:caseII}. 
As will be explained more in section \ref{Transm}, 
walls are transmuted as the relative positions are 
interchanged. 
Although the tension of individual wall changes after 
two walls pass through, the sum of tensions and the center of 
gravity are preserved. 
We have depicted the results in the middle of 
Fig.~\ref{numericalflow}.

\begin{table}
\begin{center}
\begin{tabular}{|c||c|c|}
\hline
wall & tension & position \\ \hline \hline
$2\rightarrow 1$ & $(1+n)cm$ & Re${\tau_2 / m}$                  \\
$1\rightarrow 3$ & $qcm$     & Re$\tau_1/m$                  \\
$2\rightarrow 4$ & $(q+n)cm$ & Re${n\tau_2+q\tau_1 \over (n+q)m}$ \\
$4\rightarrow 3$ & $cm$      & Re$\tau_2/m$                  \\
\hline
\end{tabular} 
\end{center}
\caption{The tensions and positions of walls in case II.}
\label{table:caseII}
\end{table}

In Case III, $m_3 > m_4 , \ m_2 +nm_4 > m_1$, 
the flow structure is as follows
\be
\left.
\begin{array}{ccccccc}
&&\alpha_4[2,4]& \rightarrow& \alpha_2[2]\\
&\nearrow&\downarrow& & \downarrow\\
\alpha_5[4]&\rightarrow&\alpha_3[0,2]& \rightarrow& \alpha_1[0]\\
\end{array} \right. .
\label{case3}
\ee
This case is very similar to Case II. 
The computed flow structure is given in the right-most 
figure of Fig.~\ref{numericalflow}.

\subsection{Wall moduli for generic hypertoric sigma models}
\label{summary}
{}From the examples that we have 
studied in detail in the preceding subsections,
we can understand the structure of BPS walls of the massive 
hypertoric sigma models.
Let us recall that a hypertoric manifold contains several
possibly noncompact
toric manifolds $M_a$ as special Lagrangian submanifolds,
and the Higgs branch of the vacua $V$ 
of the massless model is of the form 
\begin{equation}
 V=\bigcup_{a} T^*M_a .
\end{equation}
Isolated vacua of  a corresponding massive theory
sit on some of these $M_a$.
Next, we know that $\nu_A$ is zero throughout the flow 
because of the BPS equation.
Hence, BPS walls live in one of $M_a$, although $M_a$ may 
be noncompact. 
Solutions of BPS equation starting from 
such non-compact $M_a$, however, 
never reach any vacua and  have infinite tension.
Therefore they have to be discarded.
Thus, the total moduli space ${\cal M}_{\rm wall}$ of walls 
consists of the union of $M_a$'s which is compact, that is, 
\begin{equation}
 {\cal M}_{\rm wall}=\bigcup_{ K_a:\,\text{bounded}} M_a.
\end{equation}

\subsection{Attraction and repulsion between BPS walls}
\label{again}

Let us move on to study the dynamics of
BPS walls.
Using the method in Ref.~\cite{Ma},
repulsive force between two BPS walls 
was found \cite{To1} 
in the massive \hk NLSM 
on the ALE target space 
with multi-center arranged on a single line\footnote{
See also Introduction in Ref.~\cite{GTT2} and \cite{Op}. 
A similar phenomenon has been also studied in a theory with 
four SUSY \cite{Portugues:2001ah}. 
}.
As a preliminary to the later analysis, 
we first show that the similar repulsive force 
can occur in double walls interpolating 
three vacua $\alpha,\beta,\gamma $  
arranged on a straight line in the 
$\mu _A$ space.
Examples in the present paper 
are the two double wall configurations 
interpolating $\alpha _1, \alpha _3, \alpha_5$, 
and $\alpha_2,\alpha_4, \alpha _5$ 
in Fig.~\ref{toricFn} in section \ref{WLTFNLSM}. 
Interaction between two walls $\alpha\to\beta$ and 
$\beta \to \gamma$
turn out to 
be repulsive as follows. 
Each individual wall of the double wall 
is a BPS state preserving the same 1/2 supersymmetry.
Once placed side by side, however,
the configuration $\alpha\to\beta\to\gamma$ is a non-BPS state. 
This is because the sign of one of $\mu _A$'s 
must be flipped at somewhere in such a configuration 
but it is prohibited by the BPS equations 
as was explained below Eq.~(\ref{xxx}).
Unlike double walls which consist of a BPS wall and an 
anti-BPS wall, total energy density 
of such double walls  is bounded below by 
a sum of the tension of each individual BPS wall.
Therefore the total energy density must increase if we 
bring one of the BPS walls 
close to the other wall from the spatial infinity. 
We thus have shown that there exists repulsive force 
between two walls connecting vacua 
arranged on a straight line in the 
$\mu _A$ space.

The same discussion can be applied to other pairs of 
BPS walls, for instance, a set of wall 
$\alpha _4\rightarrow \alpha _3$ and 
$\alpha _3\rightarrow \alpha _1$ 
in the case I in 
subsection \ref{I}.
Namely, there exists a repulsive force between these two 
walls. 
This repulsion is explained naturally if we deform the wall 
$\alpha _ 4\rightarrow \alpha _3$ 
to the double wall configurations made of the walls 
$\alpha _4\rightarrow \alpha _5$ and 
$\alpha _5\rightarrow \alpha _3$, because 
the two walls $\alpha _5\rightarrow \alpha _3$ and 
$\alpha _3\rightarrow \alpha _1$ are the case explained 
in the last paragraph. 
Indeed, there are no walls directly going from 
$\alpha_4$ to $\alpha_1$.
In the same way we can find that there exists 
a repulsive force between 
walls $\alpha _ 2\rightarrow \alpha _4$ 
and $\alpha _ 4\rightarrow \alpha _3$. 
There turns out to be also attractive forces between 
certain sets of BPS walls 
as we will explain below. 

In the generic massive hypertoric sigma model, 
there exist many sets of vacua arranged on a straight line 
in the $\mu _A$ space as in Fig.~\ref{arrangement}.  
Thus repulsion between BPS walls is 
not special, but one of commonplace features.

Let us now study in detail the dynamics of the BPS walls 
in Case I.
Firstly, consider walls connecting $\alpha_4$ and $\alpha_3$.
{}From the flow diagram, 
we can see that the wall moduli is complex two dimensional
and that it can be identified with $W\mathbb{C}P^2{}_{1,1,n}$.
The walls can also be considered from the viewpoint of $T^*F_n$,
since both the vacua $\alpha_{3,4}$ are on $F_n$.
The Morse indices at the vacua of the flow inside $F_n$ 
are both two, as can be seen in \eqref{wow} or \eqref{case2}.
The wall moduli inside $F_n$ is of real dimension two, 
namely $\dim [S(\alpha_3) \cap U (\alpha_4)] = 2$ 
where $S$ and $U$ denote stable and unstable manifolds, 
respectively,  defined in Appendix A. 
When transversality holds this must coincides with 
$\dim S(\alpha_3) + \dim U (\alpha_4) - \dim F_n$.
However the latter is now 
$\dim S(\alpha_3) + \dim U (\alpha_4) - \dim F_n = 2+2-4 = 0
\neq 2 = \dim [S(\alpha_3) \cap U (\alpha_4)]$.
Therefore the flow violates transversality. 
According to the general discussion in section \ref{GA},
there should be walls along the cotangent direction.
We can now identify these walls with component along the 
cotangent direction as the walls inside 
$W\mathbb{C}P^2{}_{1,1,n}$, thanks to the 
embedding to the hypertoric. 

We can also see that there is another vacuum $\alpha_5$ 
outside the base $F_n$.  It may be allowed to say 
that this extra vacuum $\alpha_5$ pulled the wall along 
the cotangent. 
Note that, although $\alpha_5$ is situated at the 
coordinate infinity along the cotangent direction,
the geodesic distance from $F_n$ to $\alpha_5$ is finite.
This addition of point at `infinity' is very natural 
from the hypertoric point of view.
It is also inevitable from the viewpoint of 
the cotangent bundle of generic toric manifolds. 
This is because, as mentioned in section \ref{BPSFIT}, 
the \hk metric on $T^*M$ is unique once one fix the metric 
on $M$ and placed the condition $(*)$.
Thus, one cannot help but add $\alpha_5$ at infinity.

\begin{figure}
\begin{center}
\includegraphics[width=.4\textwidth]{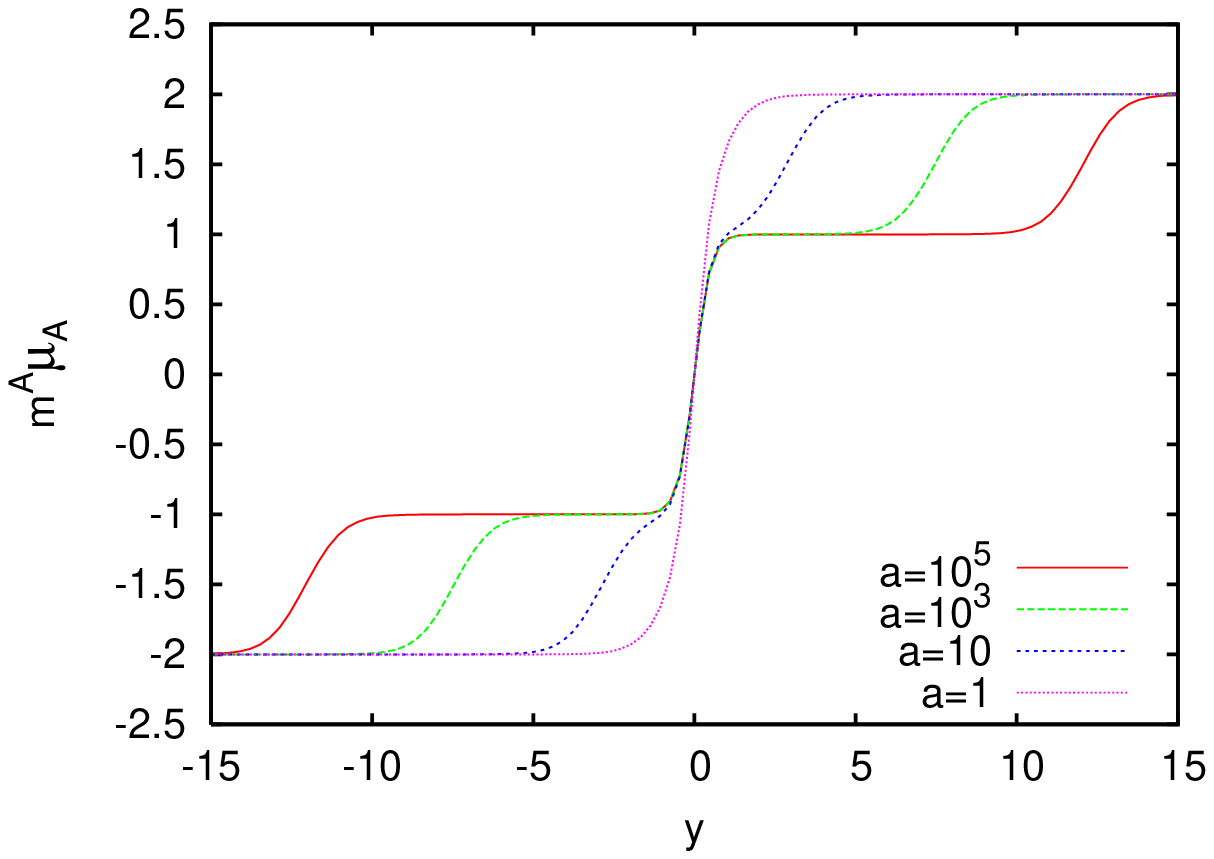}
\includegraphics[width=.5\textwidth]{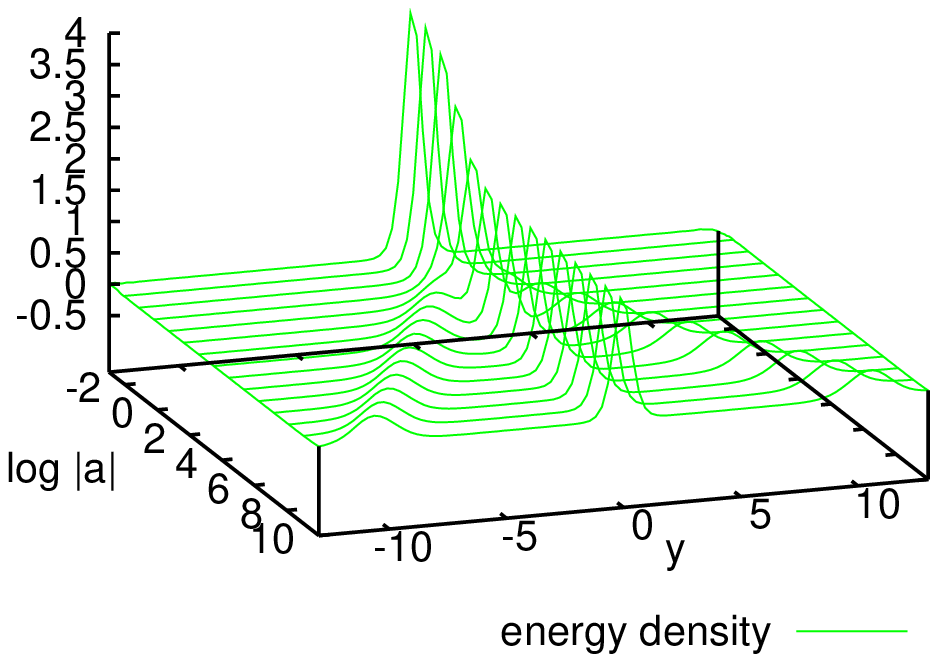}
\end{center}
\caption{
Profiles of walls $\alpha_2\to\alpha_4\to\alpha_3\to\alpha_1$.
We plotted accumulated energy density $\int_0^y \mathcal{E} dy$
in the left, and energy density $\mathcal{E}$ in the right.
\label{three}}
\end{figure}
\begin{figure}\begin{center}
\includegraphics[width=.4\textwidth]{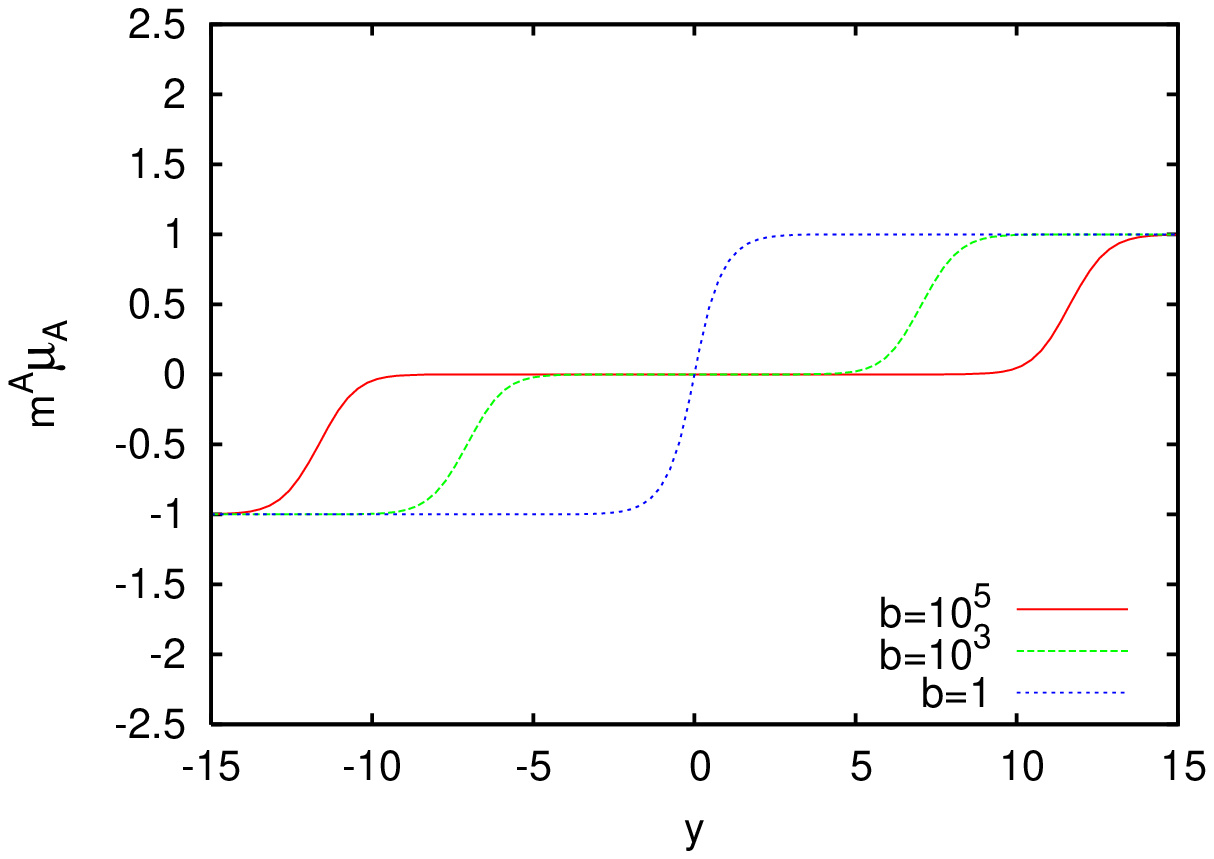}
\includegraphics[width=.5\textwidth]{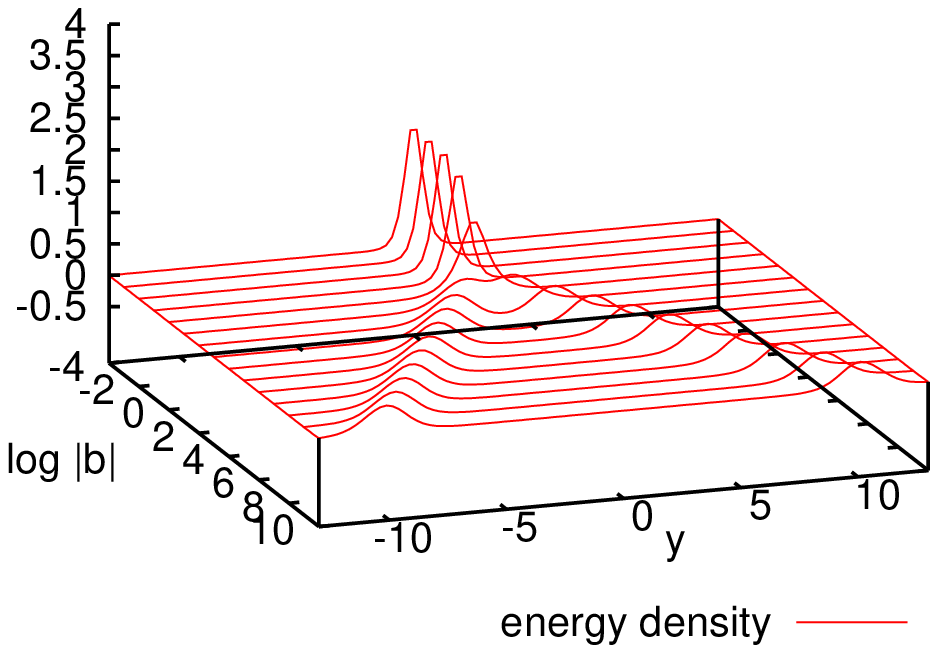}
\end{center}
\caption{Profiles of walls $\alpha_4\to\alpha_5\to\alpha_3$.
We plotted accumulated energy density 
$\int_0^y \mathcal{E} dy$ 
in the left, and energy density $\mathcal{E}$ in the right.
\label{two}}
\end{figure}

Next let us study the wall interpolating 
$\alpha_2$ and $\alpha_1$. 
In a certain limit it becomes a compressed single wall 
along $\mu_1=0$ in Fig.~\ref{toricFn}. 
The moduli space with generic moduli parameters 
is contained in $F_n$ and is of complex two dimensions. 
Contrary to this, there exist 
three BPS walls interpolating $\alpha_2\to\alpha_4$,
$\alpha_4\to\alpha_3$ and $\alpha_3\to\alpha_1$, 
and each wall carries a $U(1)$ phase. 
There exists one complex moduli parameter 
if we consider each of them independently. 
Therefore one might expect that there exist 
three complex moduli parameters for 
the configuration interpolating $\alpha_2$ and $\alpha_1$, 
whose number is not in agreement with 
the dimension of $F_n$.  
This apparent discrepancy can be resolved if we interpret 
that the wall $\alpha_2\to\alpha_1$ 
is in a sense a merger of three walls as follows.
Indeed, in a certain range of moduli, 
there are clearly three spatial regions 
where energy is concentrated as depicted in 
Fig.~\ref{three}. 
One can see there that the position of the inner wall 
is determined by the outer walls for the system to 
attain minimum energy. 
This can be understood as follows: 
First recall that 
there are repulsive force acting 
between a pair of walls 
$\alpha_2\to\alpha_4$ and $\alpha_4\to\alpha_3$ 
and between another pair 
$\alpha_4\to\alpha_3$ and $\alpha_3\to\alpha_1$, 
as discussed in the second paragraph in this subsection. 
Nevertheless there are BPS walls 
once we put in three walls. {}From these two facts 
we deduce that there must be 
attractive force between walls $\alpha_2\to\alpha_4$
and $\alpha_3\to\alpha_1$ in order to balance the 
repulsive interaction. 

\begin{figure}\begin{center}
\includegraphics[width=.8\textwidth]{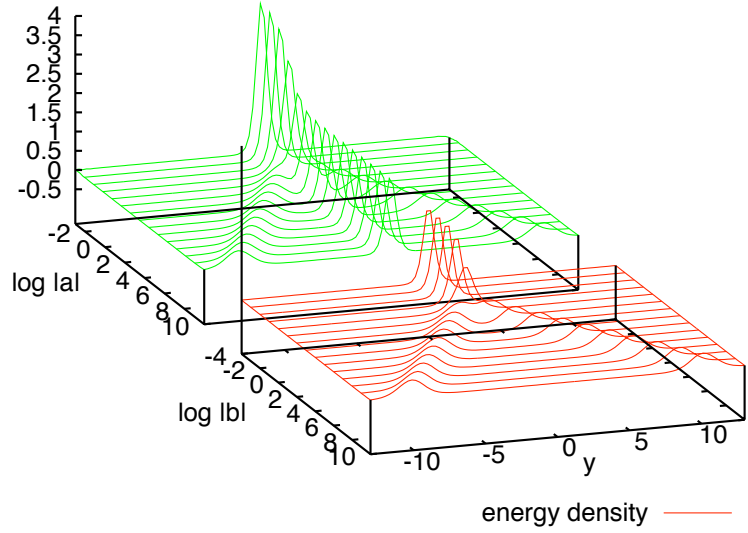}
\end{center}
\caption{Walls breaking into constituents\label{breakage}}
\end{figure}

Finally consider separating the outer two walls to 
two spatial infinities 
in the three wall system above. 
Now, the remaining wall can be 
thought of as interpolating from $\alpha_4$ to $\alpha_3$.
Although this wall is apparently a single wall system 
as exemplified in Fig.~\ref{two}, 
there are additional moduli for the cotangent direction 
and we can make it split into a double wall 
configuration consisting of walls 
$\alpha_4\to\alpha_5$ and $\alpha_5\to\alpha_3$. 
The inner walls cannot be separated in the presence of 
outer walls $\alpha_2\to\alpha_4$ and $\alpha_3\to\alpha_1$. 
This fact suggests that the repulsive force from 
the outer two walls compresses the inner two walls to 
the single wall $\alpha_4\to\alpha_3$.
The dividing process of 
a compressed single wall $\alpha_2\to\alpha_1$
to the three walls, and further dividing 
of the middle wall $\alpha_4\to\alpha_3$
to the two walls after taking off the outer two walls 
are shown in Fig.~\ref{breakage}. 

\bigskip

\begin{figure}
\centerline{\includegraphics[width=.7\textwidth]{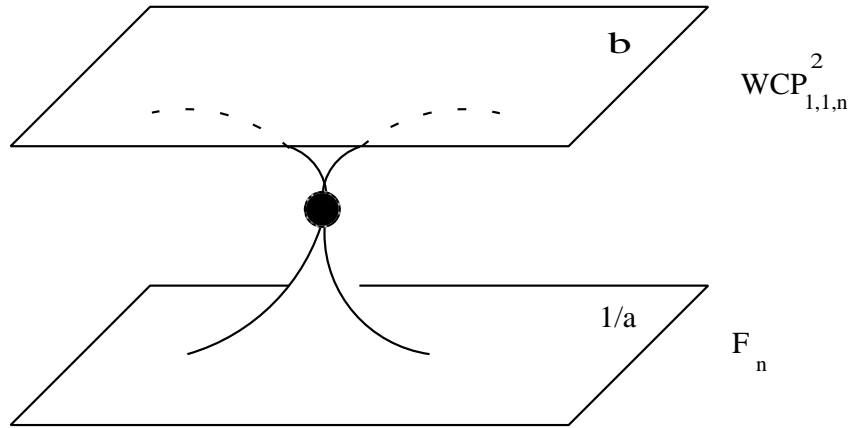}}
\caption{Schematic view of the intersection of $F_n$ and 
$W\mathbb{C}P^2{}_{1,1,n}$. 
The coordinate representing the center of mass 
is suppressed as ${\cal M}_{\rm wall}/ \mathbb{C}^*$. 
\label{intersect}}
\end{figure}

In this system, the total wall moduli is the union of 
$M_1=F_n$ and $M_2=W\mathbb{C}P^2{}_{1,1,n}$. 
Their intersection is $\mathbb{C}P^1$, which  is precisely 
the moduli space of walls $\alpha_4\to\alpha_3$, 
with the two vacua $\alpha_{3,4}$ added. 
This wall has the position $\Re \tau$ 
and the phase $\Im \tau$ associated with it 
as the only moduli. 
Let us look in more detail how $M_1$ and $M_2$ join 
at $\mathbb{C}P^1$.   We take $\tau$ and $a $ as the 
coordinates of $M_1$, and $\tau$ and $b$ as those for $M_2$.
As we have seen in \eqref{positionA} and 
\eqref{positionB}, 
$\log|a|$ and $\log|b|$ control the relative position 
of constituent walls. 
Then, the intersection is the curve $1/a=0$ in $M_1$ 
or $b=0$ in $M_2$, which is schematically illustrated 
in Fig.~\ref{intersect}. 

\begin{figure}
\centerline{\includegraphics[width=.7\textwidth]{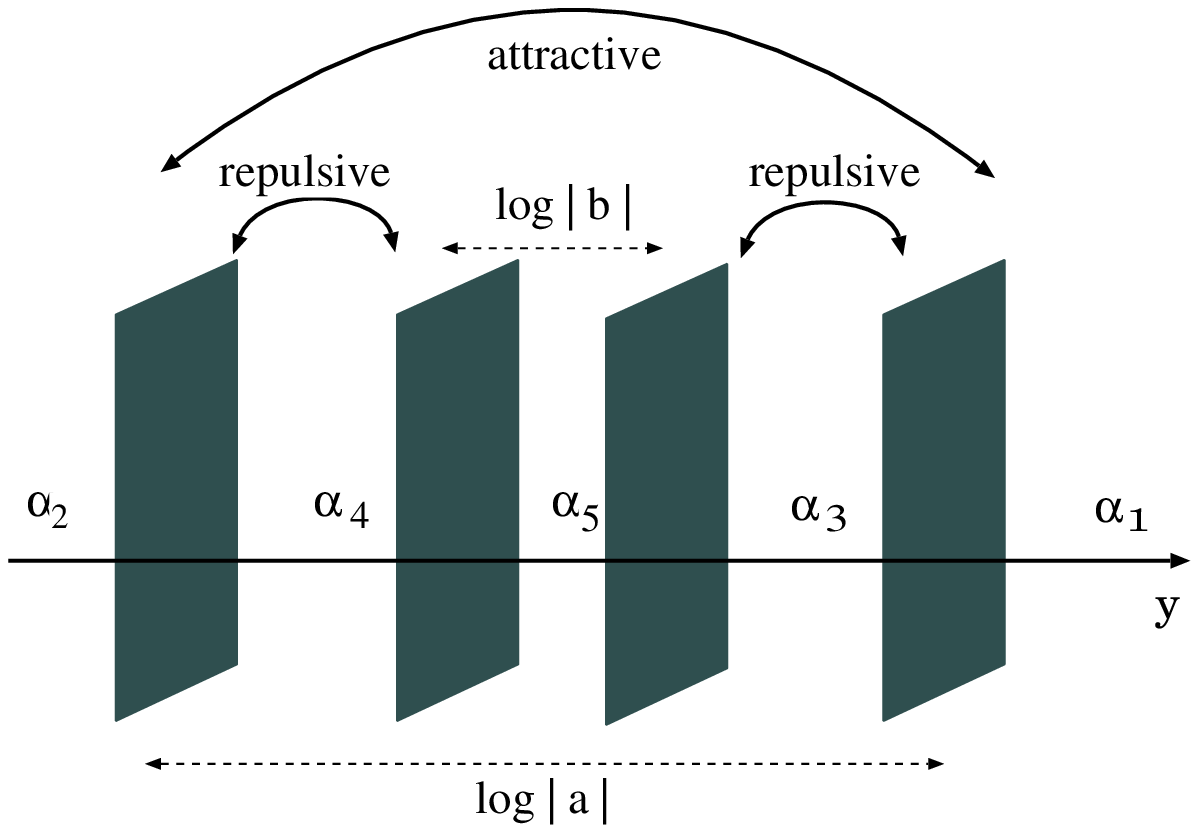}}
\caption{Schematic structure of walls interpolating from 
$\alpha_2$ to $\alpha_1$.
\label{non}}
\end{figure}

One can imagine a non-BPS wall configuration interpolating 
the all vacua successively as 
$\alpha_2\to\alpha_4\to\alpha_5\to\alpha_3\to\alpha_1$, 
as depicted in Fig.~\ref{non}.  
Only when $1/a=0$ $i.e.$ two outer walls are infinitely 
far away, BPS walls with $b\ne 0$ can appear. 
Conversely, BPS walls with $|a| < \infty$ can arise 
only when $b=0$ i.e.\ two inner walls are
completely compressed to a single wall.
In this sense, one can consider $b\cdot(1/a)=0$ as the 
defining equation of the whole BPS moduli space.

Although we used an explicit example containing $T^*F_n$ 
in order to illustrate the properties of walls, 
it is easy to see that these phenomena happen generically 
for any massive \hk sigma model with hypertoric target spaces. 
Extra vacua at infinity along cotangent direction appear 
whenever the hyperplanes bounding the toric diagram of the 
base form an obtuse angle and intersect outside of the base.
Simultaneously, there appear walls which flow outside the 
cotangent direction. 
It is extremely rare to have no obtuse angles.  
Indeed, any toric diagram with only acute or right angles 
necessarily corresponds to 
some direct product of weighted projective spaces. 
Thus, other than those cases, there are various violation 
of transversality and mutual attraction/repulsion between 
BPS walls.

\subsection{Transmutation of walls}
\label{Transm}
In the previous section
we have focused on the case I of 
the gauged linear sigma model containing $F_n$, 
and have found interesting attraction and repulsion
force between BPS walls.
Although the cases II and III do not violate in themselves 
the transversality of the flow, 
yet there are fascinating dynamics
between the BPS walls when they pass
through each other.

Recall the case of the walls in the Grassmannians \cite{INOS2}.
There, certain pair of walls cannot cross each other,
and when  the moduli parameter for 
the relative distance is formally taken to be 
negative infinity 
the walls are made into a single compressed wall.
Another pair of walls can pass through each other,
and we called them penetrable wall system.
The wall system before and after the crossing was 
equivalent, hence the walls retained their 
identities, like their tension. 

The case studied in the last subsection
is in a sense analogous to the compressed wall in Grassmannians.
Now let us study the crossing of walls in $F_n$.
It may not be surprising to the reader, who have read
all the findings discussed in the previous sections,
that we find richer dynamics in this case also.

\begin{figure}
\centerline{
\includegraphics[width=.4\textwidth]{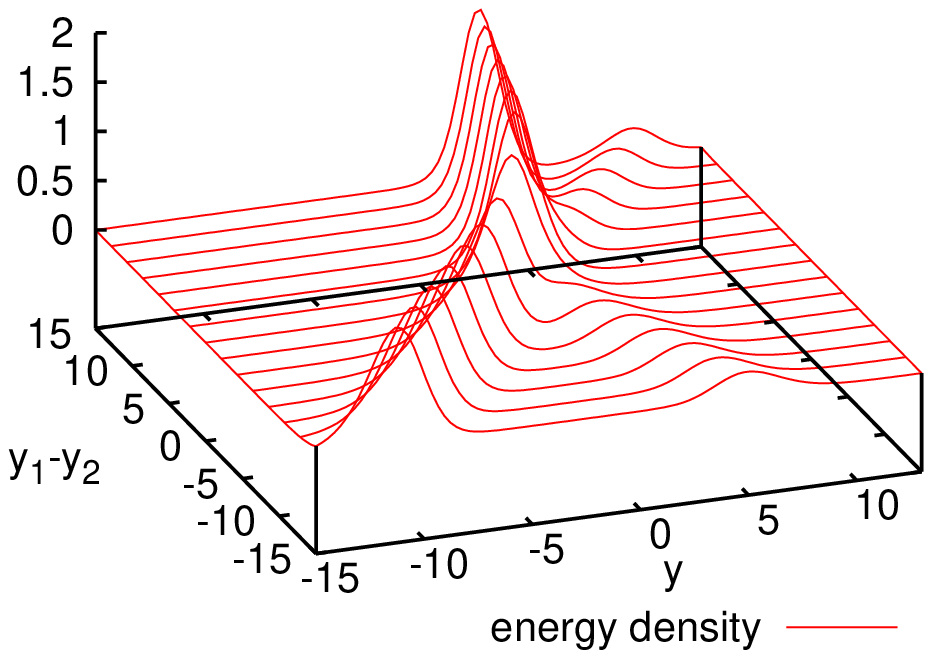}
\includegraphics[width=.4\textwidth]{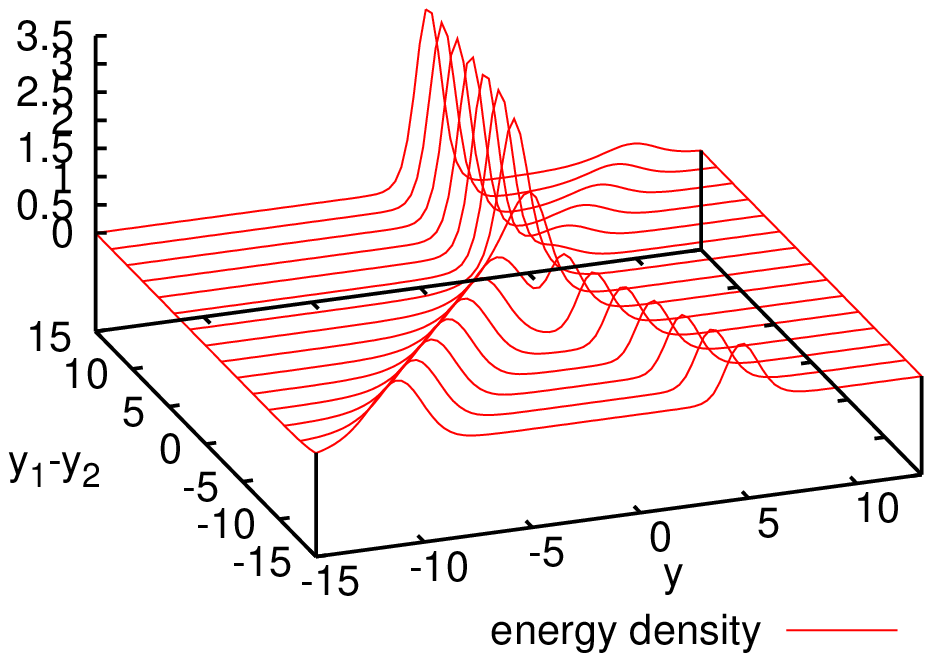}
}
\caption{Transmutation of walls when they pass through.
$m^A=(1,0,0,-1)$ in the left and
$m^A=(2,0,0,-1)$ in the right.\label{exchange}}
\end{figure}

Let us concentrate on the case II in subsection 
\ref{IIandIII} and consider walls interpolating from 
vacua $\alpha_2$ to $\alpha_3$. 
We have worked out the wall tensions and wall positions 
for the case $M_1=F_1$ and the hypermultiplet 
masses $m^A=(qm,0,0,-m)$ in section \ref{IIandIII}. 
The wall positions can be parametrized by 
$y_i \equiv {\rm Re}\tau_i/m$ using the moduli matrix 
parametrization in Eq.~(\ref{eq:moduli-matrix-II}). 
We present the result of wall positions in Fig.~\ref{exchange} 
by taking $m=1$ and $q=1$ or $q=2$ as illustrative examples.

In the limit $y_1-y_2\to\infty$, the double wall system
breaks into two walls $W_1:\alpha_2\to\alpha_4$ and
$W_2:\alpha_4\to\alpha_3$.
In the limit $y_1-y_2\to-\infty$, on the other hand,
the system breaks into $W_1':\alpha_2\to\alpha_1$ and
$W_2':\alpha_1\to\alpha_3$.
If $y_1\sim y_2$, the two walls are merged into one wall $W$.
The process can clearly be seen in Fig.~\ref{exchange}.
What distinguishes this case and the previously studied examples
of walls in Grassmannians or projective spaces is that
now the soliton before and after the encounter,
$W_1$ and $W_2'$, is not 
equivalent. $W_2$ and $W_1'$ also cannot be 
transformed to each other.  Hence, the BPS walls
transmuted through the encounter.
In the case $m^A=(1,0,0,-1)$,
the tension of the walls $W_1$ and $W_1'$ accidentally 
equals to each other as can be seen from 
Table \ref{table:caseII}.
In the case $m^A=(2,0,0,-1)$ however, the tension of walls
$W_{1,2}$ and $W_{1,2}'$ differs from each 
other completely 
with the sum of the tension unchanged.
Thus there is no doubt that the soliton has changed 
its identity.

\section{Conclusion and discussion}
\label{sc:discussion}

In this paper, we have analyzed the BPS flow equation of
massive NLSMs with eight supercharges
for various \hk target spaces.
We first have treated the most generic 
\hk manifold, and then have studied 
the case of $T^*M$ with K\"ahler manifolds $M$
and finally hypertoric manifolds, with increasing amount of detail 
in this order.

It is still difficult to survey the whole structure of
general \hk sigma models.
However, we have been 
able to see at least the connection of 
the index theorem and the problem of transversality of the 
BPS flow equation for the models with the $T^*M$ target space.
In addition, once we restricted attention to hypertoric sigma models,
we have found that the total moduli space of BPS walls, 
containing all possible topological sectors, 
is precisely the core of the hypertoric manifold, which is the union
of the special Lagrangian toric compact submanifolds
defined by the charge matrix for the hypertoric manifold.

We also have found that 
the essential reason for the non-transversality
of the flow is that the
hyperplane arrangement for $T^*F_n$ 
contains a trapezoidal cell $K_1$ with an obtuse angle, 
representing $M_1 = F_n$ in Fig.~\ref{toricFn}. 
Due to this, another vacuum $\alpha_5$ appears 
outside of the base $F_n$ and it is on another compact
special Lagrangian submanifold 
$M_2 = W{\mathbb C}P^2{}_{1,1,n}$ neighboring $F_n$.
In such a case, sets of 
vacua, 
($\alpha_2,\alpha_4,\alpha_5$) 
and ($\alpha_1,\alpha_3,\alpha_5$),  
are arranged along a straight line
in the $\mu_A$ space, and this is the cause 
of the mutual repulsion among BPS walls.

Directions of BPS flows are classified into three cases 
as in Fig.~\ref{numericalflow}, 
depending on the arrangement of the hypermultiplet masses. 
The violation of the transversality can be seen in the 
case I but not in the cases of II and III. 
We have found in the case I that, 
in the triple wall sector,  
the position of the center wall is locked 
in the center between the outer two walls 
as in Fig.~\ref{three}. 
We then have deduced 
the existence of the mutual attraction and repulsion between 
the walls as summarized in Fig.~\ref{non}. 
It would be extremely interesting to elucidate the dynamical 
mechanism leading to those forces.  
On the other hand, in the other two cases, 
the same feature of the $T^*F_n$ model, namely, 
the existence of multiple compact special Lagrangian 
submanifolds, causes completely different phenomenon, 
the transmutation of walls: 
when two walls path through, tension of one wall 
is transfered to the other with the sum of their tension 
unchanged, 
as illustrated in Fig.~\ref{exchange}. 
So we have obtained these interesting dynamics 
of walls for the $T^*F_n$ model from 
the same geometrical 
characteristic. 

The total moduli space 
is the union of $F_n$ and $W{\mathbb C}P^2{}_{1,1,n}$ 
which join together sharing the submanifold 
${\mathbb C}P^1$ as schematically illustrated 
in Fig.~\ref{intersect}. 
We would like to emphasize that we never see the 
global structure of the moduli space like this 
if we restrict ourselves only to each topological sector.

We expect that similar analysis can be extended to more 
general and complicated hypertoric sigma models. 
If the model contains $n$-polygons ($n\geq 4$) in the 
hyperplane arrangement, the non-transversality and 
transmutation may occur to walls connecting vertices of 
one of the polygons.

One of interesting future directions of research will be 
the study on the dynamics of the wall moduli. 
As in the work by Atiyah and Hitchin \cite{AtiyahHitchin} 
on the scattering of BPS monopoles, we should be able to 
study the scattering of BPS walls with each other 
because the K\"ahler potential of the moduli space 
can be obtained easily as done in 
Ref.~\cite{INOS2} in the case of Grassmannian. 
This will be particularly interesting 
when the transmutation of walls occur. 
Extending the discussion to the quantum case will be also 
worthwhile. 
Since the worldvolume theory has only four supercharges, 
there will be non-perturbative superpotential generated 
by BPS solitons of the effective theory. 
These solitons are precisely the vortices of the original 
theory, and their moduli space has been already 
analyzed in \cite{INOS3}. 
Hence, the calculation \`a la Affleck, Dine and Seiberg 
\cite{ADS} will in principle be possible.   
We can expect  richer 
non-perturbative dynamics of the BPS walls.

BPS walls in the case of Grassmannian NLSM 
can be realized using kinky D$p$-brane configuration 
in the background D($p+4$)-branes on the ALE 
space~\cite{Eto:2004vy}. 
That D-brane picture was extremely useful to understand 
various dynamics of walls. 
Therefore D-brane construction for 
the models discussed in the present paper 
is desirable. 
In particular it would be interesting to understand 
complicated dynamics of walls like the wall locking or 
the transmutation found in the present paper.

Finally, we would like to recall that much has been gained 
from taking the Killing potential of the holomorphic isometry 
as the Morse function. 
Combined with the restriction which comes from the eight 
supercharges, we should be able to extract more geometrical 
information from the BPS flow structure 
on hyper-K\"ahler target spaces. 
We are planning to investigate these problems further. 

\section*{Acknowledgement}
This work is supported in part by Grant-in-Aid for Scientific 
Research from the Ministry of Education, Culture, Sports, 
Science and Technology, Japan No.13640269 (NS) 
and 16028203 for the priority area ``origin of mass'' 
(NS). 
The works of K.Ohashi and M.N. are 
supported by Japan Society for the Promotion 
of Science (JSPS) under the Post-doctoral Research Program 
while the work of Y.T.~is by the JSPS DC1 program.
The works of Y.I.~and M.E. are supported by
 a 21st Century COE Program at 
Tokyo Tech ``Nanometer-Scale Quantum Physics'' by the 
Ministry of Education, Culture, Sports, Science 
and Technology.  
M. E. also gratefully acknowledges support from the Iwanami Fujukai Foundation.
K.~Ohta is supported in part by 
Special Postdoctoral Researchers Program at RIKEN.

\appendix

\section{Transversality and the Morse-Smale condition}
\label{MSC2}
In this appendix we collect facts about  the Morse-Smale condition
and transversality.

\subsection{Stable and unstable manifolds}
\label{SUM}
Let $M$ be a compact manifold with dimension $n$, 
$f$ be a Morse function and $\alpha, \beta, \cdots$ be 
critical points of $f$.  Let $g$ be a metric on $M$.
Define the Morse flow $\varphi_y (p)$ with 
parameter $y$ with respect to $f$ as the solution of the
differential equation \begin{equation}
\frac{d}{dy}\varphi_y(p)= g^{ij}\partial_j f(\varphi_y(p))
\end{equation}with the initial condition $\varphi_0(p)=p$.
Then, stable and unstable manifolds of $\alpha$ are defined 
as follows:
\be
S(\alpha) &=& 
\{ p\in M \bigm| \lim_{y\to +\infty} \varphi_y (p)=\alpha \}, 
\nn\\
U(\alpha) &=& 
\{ p\in M \bigm| \lim_{y\to -\infty} \varphi_y (p)=\alpha \},
\ee
respectively.
These $S(\alpha)$ and $ U(\alpha)$ are submanifolds of $M$ 
and homeomorphic to ${\mathbb{R}}^{n-n_\alpha}$ and 
${\mathbb{R}}^{n_\alpha}$ respectively, 
and  $n_\alpha$ is called the Morse  index of $\alpha$.
If we change the sign of the Morse function from $f$ to $-f$, 
the definition of stable and unstable manifold is exchanged.
The Morse function $f$ on $M$ defines 
the following two kinds of decompositions. First one is
\be
 M=\bigcup_{\alpha} S(\alpha)=\bigcup_{\alpha} U(\alpha),
\label{decomp-1}
\ee
where $\alpha$ is summed over all critical points of $f$.
So $M$ is decomposed into a sum of disjoint cells 
$S(\alpha)$ or $U(\alpha)$. Second decomposition is
\be
 M=\bigcup_{\alpha,\beta} F(\alpha,\beta), 
\ \ F(\alpha,\beta)\equiv S(\alpha)\cap U(\beta),
\label{decomp-2}
\ee
where the sum is taken over all critical points 
$\alpha,\beta$. 
Namely, $M$ is the sum of walls which go from $\beta$ 
to $\alpha$.

\subsection{Morse-Smale transversality condition}

$S(\alpha)$ intersect transversely with $U(\beta)$ if 
and only if  the following condition is satisfied
for all critical points $\alpha,\beta$ of $f$: 
\be
T_p S(\alpha) + T_p U(\beta) 
= T_p M, \ \ \text{for all\ } p\in S(\alpha)\cap U(\beta) . 
\ee
This condition is called the Morse-Smale condition 
and then $f$ is called the Morse-Smale function on $M$. 
In this case, since $S(\alpha)\cap U(\beta)$ is submanifold 
of $M$, 
\be
{\rm dim}[S(\alpha)\cap U(\beta)] 
= {\rm dim}S(\alpha) + {\rm dim}U(\beta) - {\rm dim} M,
\ee
so that we find
\be
{\rm dim}F(\alpha,\beta)
=(n-n_\alpha) + n_\beta -n =n_\beta-n_\alpha,
\ee
where $n_{\alpha,\beta}$ are indices of $\alpha,\beta$.

\section{On the absence of the vanishing theorem}
\label{sc:vanishing-th}
In the main part of the article,
we saw that the BPS flow inside $F_n$ 
violates transversality for certain range of mass parameters.
Additionally, we explicitly constructed a gauged linear sigma model
with eight supersymmetry which contains $F_n$ as a
special Lagrangian submanifold. 
Using the notation of section \ref{WLTFNLSM},
the BPS flow inside $F_n$ starting from $\alpha_4$ to $\alpha_3$ 
connects 
the vacua with the same Morse index.
As we showed in section \ref{GA}, this violation of transversality
is related to the non-vanishing of the zero modes of the operator
$\mathcal{D}_1$, which is the adjoint
of the operator $\mathcal{D}_2$
governing the deformation of the flow inside the base.

Those who also read the papers \cite{keith} and 
\cite{BPSwalls-index}, however,
may wonder why such a situation occurs, because
it was shown therein that there were no such zero modes
for gauge groups $U(1)$ and $SU(N_c)$, respectively. 
We would like to clarify why the na\"\i ve extension 
of the argument in those papers fails in our case. 
As the exposition closely follows
that of \cite{BPSwalls-index}, please place the paper 
side-by-side and compare with it.
The notation difference is tabulated in Table \ref{notation}. 
We emphasize that 
the various phenomena such as violation of transversality, 
can be regarded as one and the same 
phenomena as the absence of the vanishing theorem.

\begin{table}\[
\begin{array}{c|cccccccccccccccccccccc}
\mathrm{ours}&\partial_y & \Sigma & 
\mathcal{D}_1& \mathcal{D}_2& g^2 & c &\sigma & \eta\\
\mathrm{Ref.} \cite{BPSwalls-index}&\partial_3&\zeta^T& 
\Delta^\dagger& \Delta& e^2 & v^2 & \alpha & \beta
\end{array}\]
\caption{Comparison of notations with 
Ref.\cite{BPSwalls-index}\label{notation}}
\end{table}

Let us take a massive Abelian gauged linear sigma model 
with several $U(1)$ factor groups. 
Since the gauge coupling $g_I$ can be absorbed into the 
redefinition of the hypermultiplet charges $q_I^A$, 
we will choose to use $2g_I^2=1$. 
The equations for the fluctuations around the BPS 
wall solution can be obtained by perturbing the BPS 
equations following the argument in \cite{BPSwalls-index}. 
Denoting the fluctuations of $\Sigma^I+iW_y^I$, $h_A$ 
 and $\tilde h^A$ as $\sigma^I$, $\eta_A$ and 
$\tilde \eta^A$, respectively, we find after using the 
Gauss law and fixing the gauge 
\begin{align}
-\partial_y \sigma^I 
-\sum_A q_I^A (h^{A\dagger} \eta_A-\tilde h_A^\dagger \tilde \eta^A) 
&=0 & (\text{for each $I$}),
\label{eq:sigma-fluc}\\
(-\partial_y -\sum_I q_I^A \Sigma^I + m^A)\eta_A 
-\sum_I q_I^A h_{A} \sigma^I 
&=0 
& (\text{for each $A$}),
\label{eq:h-fluc}\\
(-\partial_y +\sum_I q_I^A \Sigma^I - m^A)\tilde \eta^A 
+\sum_I q_I^A \tilde h^{A} \sigma^I 
&=0
& (\text{for each $A$}).
\label{eq:tilde-h-fluc}
\end{align} 

As shown in section \ref{GA}, we should consider a special 
Lagrangian submanifold to obtain a nontrivial index. 
Here let us take a flow on the special Lagrangian submanifold 
$M$ defined by $\tilde h^A=0$. 
By disregarding 
the fluctuation $\tilde \eta^A$ and setting $\tilde h^A=0$ in 
Eqs.(\ref{eq:sigma-fluc})--(\ref{eq:tilde-h-fluc}), 
we obtain the operator $\mathcal{D}_2$ in \eqref{on deformation} 
as 
\begin{equation}
\mathcal{D}_2=\begin{pmatrix}
-\partial_y \unitmatrix & -q_I^A h_A^\dagger \\
 -q_I^A h_A & \mathrm{diag}(-\partial_y -\sum_I q_I^A \Sigma^I + m^A )
\end{pmatrix}  ,
\end{equation} 
acting on a vector $(\sigma^I, \eta_A)^T$. 
Here $\unitmatrix$ is the unit matrix acting on the color 
indices, and $q_I^A h_A$ should be thought of as a 
$N\times N_F$ matrix. 
The operator $\mathcal{D}_1$ is obtained as the adjoint of 
the operator $\mathcal{D}_2$ as 
 \begin{equation}
\mathcal{D}_1=\begin{pmatrix}
\partial_y \unitmatrix & -q_I^A h_A^\dagger\\
-q_I^A h_A & \mathrm{diag}(\partial_y -\sum_I q_I^A\Sigma^I + m^A )
\end{pmatrix}  .
\end{equation}

Let us try to show that there is no zero modes of the 
operator $\mathcal{D}_1$, 
imitating the argument in \cite{BPSwalls-index}.
Then we have 
\begin{multline}
\left|\mathcal{D}_1\begin{pmatrix}\sigma^I\\ \eta_A
\end{pmatrix}\right|^2
=| \partial_y \sigma^I| ^2 
+ |\sum_A q_I^A h_A^\dagger \eta_A | ^2\\
+|\partial_y\eta_A-\sum_I\eta_A q_I^A \Sigma^I +\eta_A m^A|^2
+|\sum_I q_I^A h_A \sigma_I|^2 + \text{X-terms},
\end{multline}
and the X-terms vanish when one uses the BPS equation, 
just as in \cite{BPSwalls-index}. 
Hence, the following four conditions are necessary and 
sufficient for $(\sigma^I,\eta_A)^T$ to be a zero mode:
\begin{align}
\partial_y \sigma^I&=0 &(\text{for each $I$}),
\label{eq:zero-mode1} \\
\sum_I q_I^A h_A \sigma_I& =0 &
(\text{for each $A$}),
\label{eq:zero-mode2} \\
\sum_A q^A_I h_A^\dagger \eta_A&=0 &(\text{for each $I$}),
\label{eq:zero-mode3} \\
\partial_y\eta_A-\sum_I\eta_A q_I^A \Sigma^I +\eta_A m^A&=0&
(\text{for each $A$}).
\label{eq:zero-mode4} 
\end{align}
Eqs.(\ref{eq:zero-mode1}) and (\ref{eq:zero-mode2}) 
immediately dictate the fluctuation of vector multiplet 
scalar to vanish $\sigma^I=0$. 
Let us now concentrate on Eqs.(\ref{eq:zero-mode3}) and 
(\ref{eq:zero-mode4}) to examine 
if the fluctuation $\eta_A$ of the hypermultiplet 
scalar $h_A$ has zero mode or not. 
The `moduli matrix' $q_I^A h_A$ is just a row vector, and 
the condition \eqref{eq:zero-mode3} imposes only $N$ 
conditions on the $N_f$ variables \footnote{
A similar analysis in Ref.\cite{keith} contains an incorrect 
result in its Eq.(15) with respect to this point. 
Therefore the proof of the vanishing theorem in that case 
also requires the argument that we give in this Appendix. 
}. Hence Eq.(\ref{eq:zero-mode3}) is not sufficient to 
conclude that $\eta$ vanishes. 
The equation \eqref{eq:zero-mode3} is just the condition 
\eqref{WW} which signifies that $\eta$ should be cotangent 
to the base manifold. 
Thus, we found again that the adjoint of the operator 
governing the deformation of the BPS flow inside the base 
describes the BPS flow along the cotangent direction. 

We can do better in sorting out the zero modes of 
$\mathcal{D}_1$. 
It is important to realize that the remaining zero mode 
equation (\ref{eq:zero-mode4}) for each $A$ is identical 
to the equation for the fluctuation $\tilde \eta^A$ of 
the cotangent direction $\tilde h^A$ in 
Eq.(\ref{eq:tilde-h-fluc}), 
provided we can neglect the background $\tilde h^A=0$ or 
the fluctuation of vector multiplet scalar $\sigma^I=0$. 
This again shows that the vanishing theorem (no zero mode 
of $\mathcal{D}_1$) is the same condition as the absence 
of the zero mode in the cotangent direction, even for 
finite gauge coupling rather than the NLSM (strong gauge 
coupling) limit. 
Eq.~(\ref{eq:zero-mode4}) combined with the BPS equation 
\eqref{hidebu} implies that 
the quantities $h_A\tilde \eta^A$ (no summation on $A$) 
are constant along the flow. 
Since $h_A\tilde \eta^A$ at $y\to\pm\infty$ are constrained 
to vanish, $h_A\tilde \eta^A$ (no summation on $A$) 
should always be zero, 
similarly to $\nu_A$ as noted already in \eqref{xxx}. 
Thus, when $h_A$ is nonzero for a particular flavor $A$, 
there is no zero mode $\tilde \eta^A$ along the 
corresponding $\tilde h^A$ direction. 
Let us note that the condition $h_A$ vanishes is 
equivalent to the condition $\mu_A=h_A h_A^\dagger$ 
vanishes, and recall that the hyperplanes $\mu_A=0$ 
define the boundary of the cell corresponding to the 
special Lagrangian submanifold we are considering.
Therefore, we can conclude that the vanishing theorem 
can be established when the flow is \textit{properly inside} 
the cell, and the zero mode along the cotangent direction is 
possible only for the flow 
\textit{along the boundary} of the cell.

When there is indeed a nontrivial zero mode along 
$\tilde h^A$, the same condition 
$h_A\tilde \eta^A=0$ (no summation on $A$) now 
implies $h_A=0$ for that particular $A$. 
This means that the BPS flow  in this particular topological 
sector lost one  moduli inside the base.
Additionally, the flow with nonzero $\tilde h^A$ can be 
naturally considered as the flow inside the other special 
Lagrangian submanifold $M'$ which shares the hyperplane 
$\mu_A=0$. 
Thus, for the flow inside the hypertorics, 
the appearance of the zero mode along 
the cotangent direction in a topological sector 
can be better understood as the 
existence of another compact special Lagrangian submanifold 
$M'$ sharing a boundary  with $M$.
Then, the wall in the particular topological sector is 
better interpreted as a result of taking the limit 
$\tilde h^A\to 0$ for walls inside $M'$. 
Indeed, if we choose to consider the index theorem with 
respect to the special Lagrangian submanifold $M'$ instead 
of $M$, we have a valid vanishing 
theorem and no violation of transversality. 
This point is very well illustrated in the discussion 
in section \ref{again}.
There, the wall connecting $\alpha_4$ to $\alpha_3$ 
violated transversality 
and looked like a single wall  from the viewpoint of $F_n$.
The wall is, however, decomposable further into two walls
$\alpha_4\to\alpha_5$ and $\alpha_5\to\alpha_3$ from the 
viewpoint of $W\mathbb{C}P^2{}_{1,1,n}$. 
This example strengthens our interpretation that the walls 
violating transversality should be thought of as a limit 
of the wall on another special Lagrangian
submanifold. 

We have shown in section \ref{sc:general-hk} that Morse 
index for the entire hyper-K\"ahler manifold is always $2n$ 
and does not give informations on the possible number of 
zero modes (wall directions). 
These NLSM's may be obtained by taking the strong gauge coupling 
limit from our gauged linear sigma model. 
We can obtain the corresponding result at finite gauge 
couplings by extending our analysis of zero modes as 
follows. 
Suppose we apply the same type of analysis of zero modes 
for index theorems by choosing the fluctuations 
$\sigma^I, \eta_A, \tilde \eta^A$ of the 
entire gauged linear sigma model. 
The operator $\mathcal{D}_2$ in \eqref{index theorem1} can 
be read off from 
Eqs.~(\ref{eq:zero-mode1})-(\ref{eq:zero-mode4}). 
The adjoint operator $\mathcal{D}_1$ can be easily 
worked out to give the same operator as $\mathcal{D}_2$ 
except that the sign of $\partial_y$ is just reversed. 
If we apply the same zero mode analysis as above for this 
case, we first find that vector multiplet scalar fluctuations 
should vanish. 
Then the remaining zero mode equations for $\mathcal{D}_2$ 
become 
$\partial_y\eta_A-\sum_I\eta_A q_I^A \Sigma^I +\eta_A m^A=0$ 
and 
$\partial_y\tilde \eta_A+\sum_I\tilde \eta_A q_I^A \Sigma^I 
-\tilde \eta_A m^A=0$. 
It is easy to see that these equations are identical to 
the corresponding zero mode equations for the operator 
$\mathcal{D}_1$, since the two equations are interchanged under 
$\partial_y \leftrightarrow -\partial_y$. 
This result implies that the index in Eq.~(\ref{index theorem1}) 
always vanishes if it is defined for the entire fluctuations 
of the gauged linear sigma model.


\end{document}